\documentclass[letterpaper,twocolumn,10pt]{article}
\usepackage{usenix}
\usepackage{tikz}
\usepackage{filecontents}
\usepackage{xspace} 
\usepackage{cite}
\usepackage{amsmath,amssymb,amsfonts}
\usepackage{graphicx}
\usepackage{textcomp}
\usepackage{xcolor}
\usepackage{todonotes}
\usepackage{mathtools}
\usepackage{algorithm}
\usepackage[noend]{algpseudocode}
\usepackage{subfig}
\usepackage{tabularx}
\usepackage{booktabs}
\usepackage{makecell}
\usepackage{multirow}
\usepackage{comment}
\usepackage[many]{tcolorbox}

\algnewcommand\algorithmicforeach{\textbf{for each}}
\algdef{S}[FOR]{ForEach}[1]{\algorithmicforeach\ #1\ \algorithmicdo}

\newtcolorbox{textbox}{
    boxrule = 0pt,  %
    fontupper = \small
}

\usepackage{cleveref}
\usepackage{xcolor,pifont}
	
\usepackage{color, colortbl}
\usepackage{placeins}
\usepackage{stfloats}

\newcommand{\ournameNoSpace}{GoodVibe} 
\newcommand{\ourname}{\ournameNoSpace\xspace}

\usepackage{listings}
\definecolor{codegreen}{rgb}{0,0.6,0}
\definecolor{codegray}{rgb}{0.5,0.5,0.5}
\definecolor{codepurple}{rgb}{0.58,0,0.82}
\definecolor{backcolour}{rgb}{0.95,0.95,0.92}

\let\othelstnumber=\thelstnumber
\def\createlinenumber#1#2{
    \edef\thelstnumber{%
        \unexpanded{%
            \ifnum#1=\value{lstnumber}\relax
              #2%
            \else}%
        \expandafter\unexpanded\expandafter{\thelstnumber\othelstnumber\fi}%
    }
    \ifx\othelstnumber=\relax\else
      \let\othelstnumber\relax
    \fi
}

\lstdefinestyle{customc}{
  belowcaptionskip=1\baselineskip,
  breaklines=true,
  frame=single,
  xleftmargin=0.35cm,
  xrightmargin=0.15cm,
  numbers=left,
  numbersep=5pt,  
  language=C,
  showstringspaces=false,
  basicstyle=\footnotesize\ttfamily,
  keywordstyle=\bfseries\color{green!40!black},
  commentstyle=\itshape\color{purple!40!black},
  identifierstyle=\color{blue},
  stringstyle=\color{orange},
}

\lstdefinestyle{customcArianeExploit1}{
  breaklines=true,
  frame=single,
  xleftmargin=0.4cm,
  xrightmargin=0.2cm,
  numbers=left,
  numbersep=5pt,  
  language=C,
  showstringspaces=false,
  basicstyle=\footnotesize\ttfamily,
  keywordstyle=\bfseries\color{green!40!black},
  commentstyle=\itshape\color{purple!60!black},
  identifierstyle=\color{blue},
  stringstyle=\color{yellow!50!black},
  morekeywords={asm},
  keywordstyle=[2]\bfseries\color{brown!60!black},
}

\lstdefinestyle{customcArianeExploit}{
  breaklines=true,
  frame=single,
  xleftmargin=0.4cm,
  xrightmargin=0.2cm,
  numbers=left,
  numbersep=5pt,  
  language=C,
  showstringspaces=false,
  basicstyle=\footnotesize\ttfamily,
  keywordstyle=\bfseries\color{blue},
  commentstyle=\itshape\color{green!50!black},
  identifierstyle=\color{black},
  stringstyle=\color{brown},
  morekeywords={asm},
  keywordstyle=[2]\bfseries\color{black},
}

\lstdefinestyle{customlog}{
  breaklines=true,
  frame=single,
  xleftmargin=0.35cm,
  xrightmargin=0.15cm,
  numbers=left,
  numbersep=5pt,  
  language=C,
  showstringspaces=false,
  basicstyle=\footnotesize\ttfamily,
  keywordstyle=\color{blue},
  commentstyle=\itshape\color{purple!40!black},
  identifierstyle=\color{blue},
  stringstyle=\color{orange},
  keywords=[2]{INFO},
  keywords=[3]{ERROR},x
  keywordstyle=[2]\bfseries\color{green!40!black},
  keywordstyle=[3]\bfseries\color{red!500!black},
}

\definecolor{verilogcommentcolor}{RGB}{104,180,104}
\definecolor{verilogkeywordcolor}{RGB}{49,49,255}
\definecolor{verilogsystemcolor}{RGB}{128,0,255}
\definecolor{verilognumbercolor}{RGB}{255,143,102}
\definecolor{verilogstringcolor}{RGB}{160,160,160}
\definecolor{verilogdefinecolor}{RGB}{128,64,0}
\definecolor{verilogoperatorcolor}{RGB}{0,0,128}
\definecolor{pointcolor}{RGB}{192,0,0} %
\lstdefinestyle{prettyverilog}{
   language           = Verilog,
   commentstyle       = \color{verilogcommentcolor},
   alsoletter         = \$'0123456789\`,
   literate           = *{+}{{\verilogColorOperator{+}}}{1}%
                         {-}{{\verilogColorOperator{-}}}{1}%
                         {@}{{\verilogColorOperator{@}}}{1}%
                         {;}{{\verilogColorOperator{;}}}{1}%
                         {*}{{\verilogColorOperator{*}}}{1}%
                         {?}{{\verilogColorOperator{? }}}{1}%
                         {:}{{\verilogColorOperator{:}}}{1}%
                         {<}{{\verilogColorOperator{<}}}{1}%
                         {>}{{\verilogColorOperator{> }}}{1}%
                         {!}{{\verilogColorOperator{!}}}{1}%
                         {^}{{\verilogColorOperator{^}}}{1}%
                         {|}{{\verilogColorOperator{|}}}{1}%
                         {||}{{\verilogColorOperator{|| }}}{1}%
                         {=}{{\verilogColorOperator{= }}}{1}%
                         {==}{{\verilogColorOperator{== }}}{1}%
                         {=>}{{\verilogColorOperator{=> }}}{1}%
                         {[}{{\verilogColorOperator{[}}}{1}%
                         {]}{{\verilogColorOperator{]}}}{1}%
                         {(}{{\verilogColorOperator{(}}}{1}%
                         {)}{{\verilogColorOperator{)}}}{1}%
                         {,}{{\verilogColorOperator{,}}}{1}%
                         {.}{{\verilogColorOperator{.}}}{1}%
                         {~}{{\verilogColorOperator{$\sim$}}}{1}%
                         {\%}{{\verilogColorOperator{\%}}}{1}%
                         {\&}{{\verilogColorOperator{\& }}}{1}%
                         {\&\&}{{\verilogColorOperator{\&\& }}}{1}%
                         {\#}{{\verilogColorOperator{\#}}}{1}%
                         {\ /\ }{{\verilogColorOperator{\ /\ }}}{3}%
                         {\ _}{\ \_}{2}%
                        ,
   morestring         = [s][\color{verilogstringcolor}]{"}{"},%
   identifierstyle    = \color{black},
   vlogdefinestyle    = \color{verilogdefinecolor},
   vlogconstantstyle  = \color{verilognumbercolor},
   vlogsystemstyle    = \color{verilogsystemcolor},
   basicstyle         = \scriptsize\fontencoding{T1}\ttfamily,
  columns=fullflexible, 
   keywordstyle       = \bfseries\color{verilogkeywordcolor},
   morekeywords      = {val, when, port, coverage, unique},
   numbers            = left,
   numbersep          = 5pt,
   tabsize            = 2,
   escapeinside       = {/*!}{!*/},
   upquote            = true,
   sensitive          = true,
   showstringspaces   = false, %
   frame              = single,
   breaklines         = true,
   abovecaptionskip   = 0pt,
   belowcaptionskip   = 2pt,   
   xleftmargin        =0.35cm,
   xrightmargin       =0.15cm,
   captionpos         = t,
   emph               = {Point, Point0, Point1, Point2, Point3, Point4, Point5, Point6, Point7, Point8, Point9},
   emphstyle          =\color{pointcolor},%
   emph               = {[2] STVEC,SCOUNTEREN,MSTATUS,MTVEC,ML1_ICACHE_MISS,ML1_DCACHE_MISS,MITLB_MISS,MDTLB_MISS,
                             MLOAD,MSTORE,MEXCEPTION,MEXCEPTION_RET,MBRANCH_JUMP,MCALL,MRET,MMIS_PREDICT,MSB_FULL,
                             MIF_EMPTY,MHPM_COUNTER_17,MHPM_COUNTER_18,MHPM_COUNTER_19,MHPM_COUNTER_20,MHPM_COUNTER_21,
                             MHPM_COUNTER_22,MHPM_COUNTER_23,MHPM_COUNTER_24,MHPM_COUNTER_25,MHPM_COUNTER_26,MHPM_COUNTER_27,
                             MHPM_COUNTER_28,MHPM_COUNTER_29,MHPM_COUNTER_30,MHPM_COUNTER_31,property,endproperty, s_eventually}, %
   emphstyle          = {[2]\bfseries\color{verilogkeywordcolor}}
}

\makeatletter

\newcommand\language@verilog{Verilog}
\expandafter\lst@NormedDef\expandafter\languageNormedDefd@verilog%
  \expandafter{\language@verilog}
  
\lst@SaveOutputDef{`'}\quotesngl@verilog
\lst@SaveOutputDef{``}\backtick@verilog
\lst@SaveOutputDef{`\$}\dollar@verilog

\newcommand\getfirstchar@verilog{}
\newcommand\getfirstchar@@verilog{}
\newcommand\firstchar@verilog{}
\def\getfirstchar@verilog#1{\getfirstchar@@verilog#1\relax}
\def\getfirstchar@@verilog#1#2\relax{\def\firstchar@verilog{#1}}

\newcommand\addedToOutput@verilog{}
\lst@AddToHook{Output}{\addedToOutput@verilog}

\newcommand\constantstyle@verilog{}
\lst@Key{vlogconstantstyle}\relax%
   {\def\constantstyle@verilog{#1}}

\newcommand\definestyle@verilog{}
\lst@Key{vlogdefinestyle}\relax%
   {\def\definestyle@verilog{#1}}

\newcommand\systemstyle@verilog{}
\lst@Key{vlogsystemstyle}\relax%
   {\def\systemstyle@verilog{#1}}

\newcount\currentchar@verilog
  
\newcommand\@ddedToOutput@verilog
{%
   \ifnum\lst@mode=\lst@Pmode%
      \expandafter\getfirstchar@verilog\expandafter{\the\lst@token}%
      \expandafter\ifx\firstchar@verilog\backtick@verilog
         \let\lst@thestyle\definestyle@verilog%
      \else
         \expandafter\ifx\firstchar@verilog\dollar@verilog
            \let\lst@thestyle\systemstyle@verilog%
         \else
            \expandafter\ifx\firstchar@verilog\quotesngl@verilog
               \let\lst@thestyle\constantstyle@verilog%
            \else
               \currentchar@verilog=48
               \loop
                  \expandafter\ifnum%
                  \expandafter`\firstchar@verilog=\currentchar@verilog%
                     \let\lst@thestyle\constantstyle@verilog%
                     \let\iterate\relax%
                  \fi
                  \advance\currentchar@verilog by \@ne%
                  \unless\ifnum\currentchar@verilog>57%
               \repeat%
            \fi
         \fi
      \fi
   \fi
}

\lst@AddToHook{PreInit}{%
  \ifx\lst@language\languageNormedDefd@verilog%
    \let\addedToOutput@verilog\@ddedToOutput@verilog%
  \fi
}

\newcommand{\verilogColorOperator}[1]
{%
  \ifnum\lst@mode=\lst@Pmode\relax%
   {\bfseries\textcolor{verilogoperatorcolor}{#1}}%
  \else
    #1%
  \fi
}

\makeatother

\lstdefinestyle{mystyle}{
    commentstyle=\textit,
    keywordstyle=\textbf,
    stringstyle=\color{codepurple},
    basicstyle=\ttfamily,
    breakatwhitespace=false,         
    breaklines=true,      
    frame=single, 
    framexleftmargin=\parindent,
    captionpos=b,                    
    keepspaces=true,                 
    numbers=left,    
    numberstyle=\normalsize,
    stepnumber=1,
    numbersep=5pt,   
    xleftmargin=1.5\parindent,
    showspaces=false,                
    showstringspaces=false,
    showtabs=false,                  
    tabsize=2
}

\lstdefinestyle{CStyle}{
  backgroundcolor=\color{gray!5},
  commentstyle=\color{green!70!black},
  keywordstyle=\color{magenta!70!black},
  numberstyle=\tiny\color{gray},
  stringstyle=\color{blue!70!black},
  basicstyle=\footnotesize\ttfamily,
  breakatwhitespace=false,
  breaklines=true,
  captionpos=b,
  keepspaces=true,
  numbers=left,
  numbersep=5pt,
  showspaces=false,
  showstringspaces=false,
  showtabs=false,
  tabsize=2,
  language=C
}

\begin{document}

\date{}

\title{\Large \bf GoodVibe: Security-by-Vibe for LLM-Based Code Generation}

\author{
{\rm Maximilian Thang}\\
Technical University of Darmstadt
\and
{\rm Lichao Wu}\\
University of Bristol
\and
{\rm Sasha Behrouzi}\\
Technical University of Darmstadt
\and
{\rm Mohamadreza Rostami}\\
Technical University of Darmstadt
\and
{\rm Jona te Lintelo}\\
Radboud University
\and
{\rm Stjepan Picek}\\
University of Zagreb Faculty of Electrical Engineering and Computing  \& Radboud University
\and
{\rm Ahmad-Reza Sadeghi}\\
Technical University of Darmstadt
} %

\maketitle

\begin{abstract}
Large language models (LLMs) are increasingly used for code generation in fast, informal development workflows, often referred to as vibe coding, where speed and convenience are prioritized, and security requirements are rarely made explicit. In this setting, models frequently produce functionally correct but insecure code, creating a growing security risk. Existing approaches to improving code security rely on full-parameter fine-tuning or parameter-efficient adaptations, which are either costly and prone to catastrophic forgetting or operate at coarse granularity with limited interpretability and control.

We present GoodVibe, a neuron-level framework for improving the security of code language models \emph{by default}. GoodVibe is based on the key insight that security-relevant reasoning is localized to a small subset of neurons. We identify these neurons using gradient-based attribution from a supervised security task and perform neuron-selective fine-tuning that updates only this security-critical subspace. To further reduce training cost, we introduce activation-driven neuron clustering, enabling structured updates with minimal overhead. We evaluate GoodVibe on six LLMs across security-critical programming languages, including C++, Java, Swift, and Go. GoodVibe substantially improves the security of generated code while preserving general model utility, achieving up to a 2.5× improvement over base models, achieving performance competitive with full fine-tuning while using over 4\,700× fewer trainable parameters, and reducing training computation by more than 3.6× compared to the parameter-efficient baseline (LoRA). Our results demonstrate that neuron-level optimization offers an effective and scalable approach to securing code generation without sacrificing generality.

\end{abstract}

\section{Introduction}
\label{sec:introduction}

Large language models (LLMs) have become integral to modern software development, supporting code completion, program synthesis, and automated refactoring~\cite{ozkaya2023application}. Recent code-specialized models achieve impressive functional correctness and productivity gains~\cite{chen2021evaluating,GeminiCodeAssist}, and are increasingly adopted in both professional and casual development workflows~\cite{AICodeUsage}. This informal, exploratory usage, often referred to as “vibe coding”, prioritizes speed and convenience over careful design or review, and typically proceeds without explicit security considerations~\cite{ray2025review,zhao2025vibe}. In such settings, insecure code can be generated, copied, and reused with minimal audit. As a result, security vulnerabilities introduced by LLM-generated code have emerged as a serious concern. Prior work shows that even highly capable models frequently produce insecure patterns, including injection vulnerabilities, hallucinated dependencies, and unsafe memory operations~\cite{schreiber2025security,spracklen2025we}. When incorporated into production systems, these issues can be directly exploitable and may propagate through reused components and dependencies, amplifying risk across the software supply chain.

\noindent\textbf{Secure LLM-generated Code:\ An Unsolved Problem.} A natural mitigation strategy is to fine-tune LLMs on security-focused datasets. However, full-parameter fine-tuning is computationally expensive and often fragile, with security gains frequently accompanied by catastrophic forgetting or degraded general coding performance~\cite{he2023large,he2024instruction,qi2023fine}. Parameter-efficient methods such as LoRA~\cite{hu2022lora} reduce training cost but operate at coarse granularity, offering limited interpretability or control over how security behavior is encoded~\cite{hu2022lora,ding2023parameter}. In practice, this makes it difficult to reason about robustness or ensure that security improvements generalize beyond the fine-tuning data.
Prompt engineering is another common approach, explicitly instructing models to ``write secure code''. Such methods can improve security behavior, particularly when combined with structured prompting or external harness-level safeguards. However, these approaches still depend on the presence and consistency of security-aware prompting during deployment. In prevalent vibe-coding workflows, where prompts are often short, informal, and functionality-oriented, security guidance may be underspecified or omitted entirely. Accordingly, improving security behavior by default, without sacrificing core coding capabilities or relying solely on explicit prompting strategies, remains an important challenge.

\noindent\textbf{Our Goal and Contribution.} In this work, we introduce \ourname, a neuron-level security optimization framework for code LLMs. Our central insight is that security-relevant reasoning is not uniformly distributed across model parameters, but concentrated in a small subset of neurons. Exploiting this structure enables targeted security optimization without the drawbacks of existing fine-tuning paradigms.

At a high level, \ourname proceeds in two stages. First, it identifies \emph{security neurons}, neurons that play a dominant role in security-related reasoning. This identification is non-trivial: individual neurons do not carry explicit semantic labels, and security-relevant behavior is not directly observable from activations alone. \ourname addresses this challenge by recasting security neuron identification as a supervised security evaluation problem. By prompting the model to distinguish between secure and insecure code and analyzing neuron-level gradient magnitudes from a single backward pass, we infer which neurons exert a strong influence on security-related decisions. 
These neurons define a security-critical subspace that captures the model’s internal representation of secure coding behavior.
Second, the model undergoes neuron-based optimization restricted to this security-critical subspace, while all other parameters are frozen. To further improve efficiency and stability, \ourname leverages neuron activation behavior to cluster security neurons, allowing groups of neurons with similar functional roles to share update directions during training. This design substantially reduces the effective number of trainable parameters while preserving the model’s capacity to adapt security-relevant representations.
Across six open-source models and four security-critical programming languages, \ourname significantly improves the security of generated code compared with other methods, in the meantime largely preserving functional correctness and syntactic validity. 
Our results point toward more interpretable, controllable, and robust approaches to securing LLM-based code generation, especially in security-agnostic, real-world workflows. Our contributions are as follows:

\begin{itemize}
    \item We introduce a new approach to securing code LLMs by showing that security-relevant reasoning is structurally localized within a small subset of neurons, rather than being uniformly distributed across model parameters.
    \item We propose a gradient-based method to identify security-critical neurons by recasting security analysis as a supervised evaluation task, enabling reliable neuron-level attribution without requiring explicit semantic labeling.
    \item We introduce a novel cluster-based fine-tuning mechanism by grouping security neurons with similar functional roles and updating them within clusters. This method significantly reduces the training complexity without compromising security adaptation capability.
    \item We demonstrate that \ourname significantly improves code security across multiple models and programming languages. On C++ and Java, it achieves up to a 2.5× increase in secure code generation over base models, while achieving performance competitive with full fine-tuning using over 4\,700× fewer trainable parameters. Moreover, \ourname achieves comparable or stronger performance than parameter-efficient baselines such as LoRA across the evaluated languages, including C++, Java, Swift, and Go, while incurring substantially lower computational cost.
\end{itemize}

The rest of this paper is organized as follows: Section~\ref{sec:preliminaries} reviews background concepts and related work on code LLMs and fine-tuning techniques. Section~\ref{sec:framework} introduces the \ourname framework and details its neuron-level design. Section~\ref{sec:Implementation} presents implementation details, while Section~\ref{sec:Performance Evaluation} evaluates \ourname through case studies and experiments. Section ~\ref{sec:ablation study} provides ablation analysis, and Section~\ref{sec:discussion} discusses the critical aspects of \ourname. Sections~\ref{sec:related_works} and~\ref{sec:conclusions} provide related works and a conclusion, respectively.

\section{Preliminaries}
\label{sec:preliminaries}

\subsection{Code Large Language Model}
Code LLMs are typically instantiated as transformer-based autoregressive models trained to predict the next token in a sequence of code tokens. Given an input sequence $x = (x_1, \ldots, x_n)$, the model learns a conditional distribution $p(x_{n+1} \mid x_{\le n})$ over a vocabulary that includes keywords, identifiers, operators, and literals. Pre-training is performed on large-scale code corpora spanning multiple programming languages and development contexts~\cite{roziere2023code}.

While code LLMs demonstrate strong performance in terms of functional correctness and generalization across coding tasks, they do not explicitly model security properties during pre-training. As a result, security-relevant reasoning, such as identifying unsafe data flows or vulnerable API usage, emerges implicitly from training data rather than being directly supervised. This limitation becomes particularly critical in informal development workflows, such as vibe coding, where prompts emphasize rapid functionality and security considerations are rarely stated. In such settings, the model’s implicit and weakly encoded security knowledge is often insufficient to prevent insecure defaults. Understanding how internal model components represent and influence security-related behavior is, therefore, essential for enabling targeted security adaptation that improves default generation quality without relying on explicit security prompts.

\subsection{Parameter Efficient Fine-Tuning}
Fine-tuning is a standard approach for adapting pre-trained language models to downstream tasks, including security-related objectives such as vulnerability detection or secure code generation. 
Parameter-Efficient Fine-Tuning (PEFT) methods seek to reduce this cost by restricting optimization to a subset of parameters while freezing the pre-trained backbone. Common PEFT approaches include adapter-based methods, which insert lightweight trainable modules between transformer layers; prefix or prompt tuning, which prepends trainable embeddings to model inputs; and low-rank adaptation techniques that constrain parameter updates to low-dimensional subspaces~\cite{xu2026parameter,hu2022lora}.
From an optimization perspective, PEFT methods modify model behavior by introducing or updating parameters at the level of layers or submodules. These approaches have been shown to retain much of the pre-trained model’s general capability while improving task-specific performance~\cite{zhang2025parameter,xu2026parameter}.
However, since PEFT operates at coarse granularity, it provides limited visibility into how task-relevant knowledge is encoded within the model’s internal representations, particularly at individual neurons.

\section{\ourname}
\label{sec:framework}

\subsection{Threat Model}
\label{subsec:threat model}

\noindent\textbf{Goal and Assumptions.}
Our goal is to improve the \emph{intrinsic security awareness} of code language models under normal, non-adversarial usage. We focus on the common scenario in which a developer relies on a code LLM for everyday programming tasks, such as code completion and rapid prototyping, where prompts primarily emphasize functionality and development speed, and security requirements are implicit, under-specified, or absent. 

We assume the code LLM is deployed as-is to the user after training. While the pretrained model may originate from a third party and could have been trained on imperfect or insecure data, we do not assume that an adversary has direct access to modify the model parameters during deployment. After deployment, the adversary’s interaction with the model is limited to standard prompting, just like any other user.

\noindent\textbf{Threat Addressed.}
The threat we address arises when a model operating under benign, security-agnostic prompts generates code that is functionally correct but insecure. Common examples include missing input validation when handling external data, unsafe memory management in low-level code, incomplete or incorrect authentication logic, and insecure usage of APIs that require additional safeguards. These vulnerabilities are not the result of malicious prompts or intentional misuse. Instead, they stem from the model’s learned coding behavior during pretraining on large-scale code corpora, which contain a substantial amount of insecure, incomplete, or context-dependent code. Furthermore, since code generated by LLMs is often copied, adapted, and integrated into larger codebases with minimal modification, insecure defaults can easily propagate downstream. Our objective is therefore to raise the \emph{baseline security quality} of generated code, making secure behavior more likely without explicit request.

\noindent\textbf{Out-of-Scope Scenarios.}
We do not aim to defend against active attacks on the model, such as adversarial prompting, jailbreak attempts, or the intentional generation of insecure or malicious code. These scenarios fall under the broader domain of safety alignment and misuse prevention and are orthogonal to the focus of this work. We also do not evaluate structural vulnerabilities, e.g., Python-specific issues such as insecure deserialization flows, unsafe dynamic imports, or cross-module path traversal, that require repository-level, multi-file, or deployment-context analysis.

\subsection{Design Intuition}
\label{subsec:design}
\ourname is motivated by an analogy to how humans regulate behavior. Humans possess broad, transferable knowledge, but safe behavior is enforced through learned constraints, such as safety norms or professional standards, that guide \emph{how} this knowledge is applied. These constraints do not replace general capability; they act as internal control mechanisms. We argue that a similar structure exists in code language models. A pretrained code LLM already knows how to generate syntactically correct and functionally effective code, including for security-sensitive tasks.\footnote{We refer to secure code as implementations that follow common security best practices (e.g., validating inputs and using safe memory operations), and insecure code as functionally correct code that omits such safeguards.} The core issue is not insufficient capability, but \emph{insufficient security awareness}: when security requirements are implicit or absent, the model defaults to common but potentially insecure coding patterns. In fact, improving security in this setting does not require relearning how to code. Instead, it requires reinforcing internal control mechanisms that bias generation toward secure practices. Retraining the entire model is thus unnecessary and potentially harmful. \ourname aims to identify and selectively strengthen a small set of such security control mechanisms.

Since such control mechanisms are not explicitly labeled within the model, as discussed in Section~\ref{subsec:security structure identification}, \ourname infers it indirectly by observing how the model behaves when asked to \emph{judge} code security. Neurons that exert a strong influence on security-related decisions are treated as the model’s internal security controls. By selectively adapting these neurons, \ourname increases the likelihood that security considerations are applied during code generation, even when the user does not explicitly request secure code.
Furthermore, in large transformer models, individual neurons typically capture low-level features or partial signals, while higher-level behaviors are distributed across multiple neurons~\cite{elhage2021mathematical}. Following this, we argue that the security control logic emerges from groups of neurons that respond similarly to security-relevant patterns. As detailed in Section~\ref{subsec:neuron selective tuning}, \ourname captures this structure by jointly adapting neurons with correlated behavior, enabling the model to efficiently and robustly internalize security constraints. In doing so, \ourname improves security behavior while preserving the model’s overall coding capabilities.

\begin{figure*}[t]
    \centerline{\includegraphics[width=\linewidth]{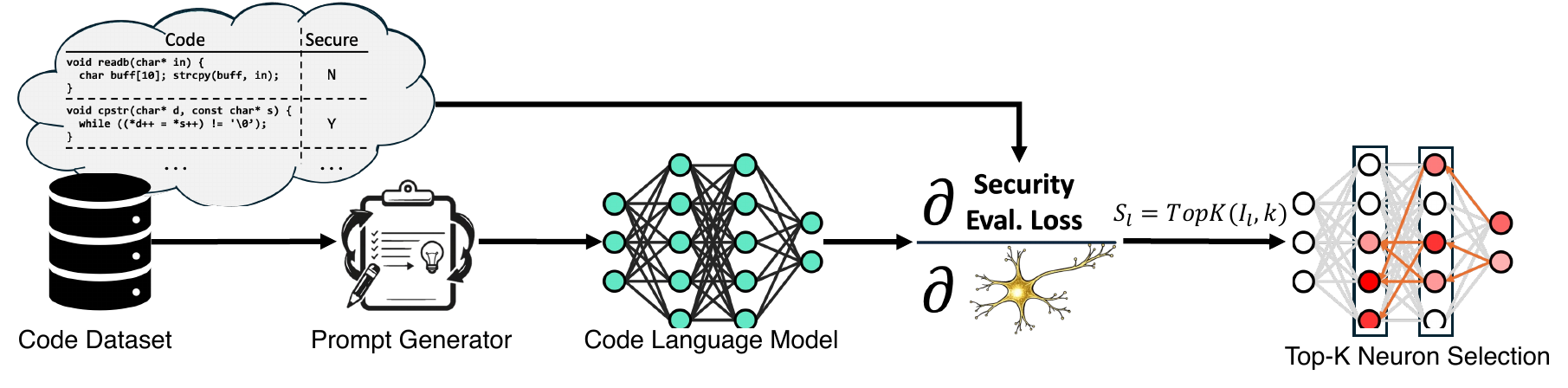}}
    \caption{An Overview of the Security Neuron Identification Pipeline.}
    \label{fig:overviewstep1}
\end{figure*}

\subsection{Security Neuron Identification}
\label{subsec:security structure identification}

\ourname first identifies the neurons responsible for security-related reasoning. 
Our goal is to find neurons whose internal representations exert a strong influence on the model’s security-related decisions.
Figure~\ref{fig:overviewstep1} presents an overview of the security neuron identification steps. Formally, let $f_\theta$ denote a pre-trained code LLM with parameters $\theta$, and let $\mathcal{D} = \{x_n, y_n\}_{n=1}^{N}$ be a dataset of code snippets $x_n$ with corresponding security labels $y_n \in \{0,1\}$, indicating whether the snippet is secure or insecure. We define a security evaluation task using a \emph{security loss function} $\mathcal{L}(f_\theta(x), y)$, which measures the model’s error when predicting the security label of a code snippet. 

To quantify the contribution of individual neurons, \ourname leverages gradient-based attribution. Gradients capture the sensitivity of the security loss to changes in model parameters and therefore provide a principled signal for identifying neurons that causally influence security-related predictions.

Consider a transformer layer $l$ with hidden dimension $d_l$. Each neuron corresponds to one dimension of the layer’s intermediate representation and is associated with a subset of parameters $\theta_{l,i}$, where $i \in \{1, \dots, d_l\}$. We define the security score (SS) of neuron $i$ in layer $l$ as the expected magnitude of the gradient of the security loss with respect to its parameters:
\begin{equation}
SS_{l,i}
=
\mathbb{E}_{(x,y)\sim\mathcal{D}}
\left[
\left|
\frac{\partial \mathcal{L}(f_\theta(x), y)}
{\partial \theta_{l,i}}
\right|
\right].
\end{equation}
In practice, this expectation is approximated by averaging gradient magnitudes across all samples in the dataset.

For each layer $l$, we select the top-$k$ neurons with the highest importance scores:
\begin{equation}
\mathcal{S}_l = \operatorname{TopK}(SS_{l,\cdot}, k).
\label{eq:top-k}
\end{equation}

We perform neuron selection independently for each layer rather than globally across the entire model, which prevents neurons from a small number of layers from dominating the security-critical subspace due to scale differences in gradient magnitudes. 
More importantly, it reflects the structure of secure code reasoning, which spans multiple levels of abstraction: low-level operations such as memory handling or input sanitization are more likely captured in earlier layers, while higher-level security logic, including API usage patterns and control-flow decisions, emerges in later layers.

Finally, the union of selected neurons across all layers defines the \emph{security-critical neuron subspace}:
\begin{equation}
\mathcal{S} = \bigcup_{l=1}^{L} \mathcal{S}_l.
\end{equation}

This subspace serves as the target for neuron-selective optimization in subsequent stages of \ourname.

\subsection{Cluster-based Security Optimization}
\label{subsec:neuron selective tuning}

\begin{figure*}[t]
    \centerline{\includegraphics[width=\linewidth]{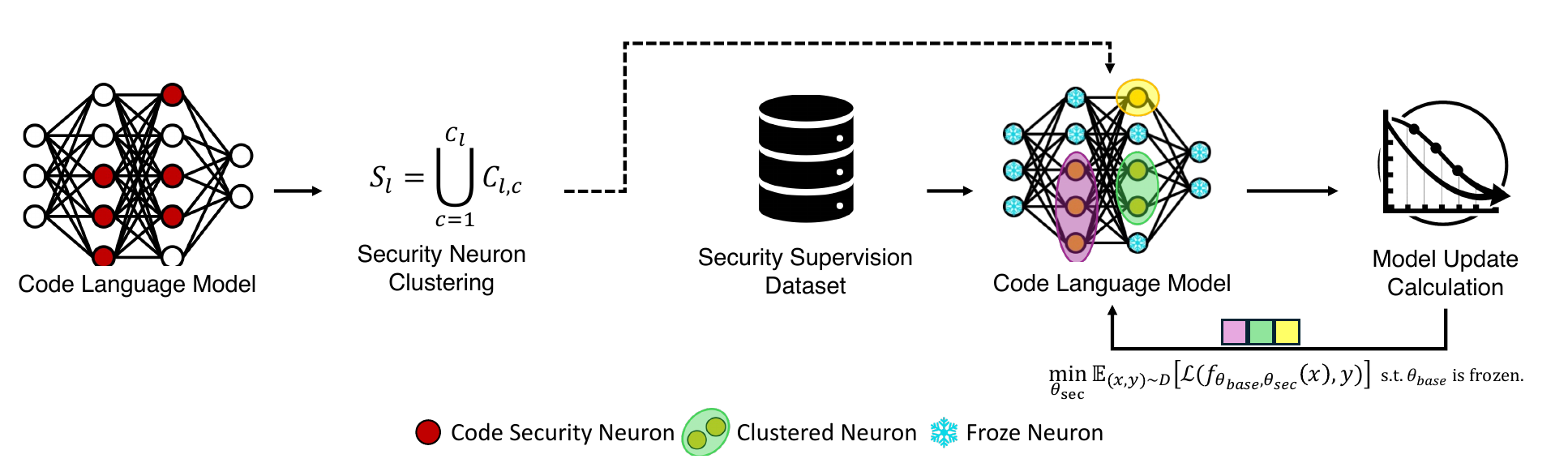}}
    \caption{An Overview of the Cluster-based Security Optimization Pipeline. }
    \label{fig:overviewstep23}
\end{figure*}

After identifying the security-critical neuron subspace, \ourname optimizes these security structures to improve the model’s security behavior while preserving its generalization capability. An overview of the \ourname security optimization pipeline is shown in Figure~\ref{fig:overviewstep23}. The key idea is to restrict parameter updates to neurons directly involved in security-related reasoning and to freeze all remaining parameters.

Formally, given a pre-trained model $f_\theta$, the full parameter set $\theta$ can be partitioned as:
\begin{equation}
    \theta = \theta_{\text{sec}} \cup \theta_{\text{base}},
\end{equation}
where $\theta_{\text{sec}}$ denotes parameters associated with the security neurons identified in Section~\ref{subsec:security structure identification}, and $\theta_{\text{base}}$ contains all remaining parameters.

Given a security supervision dataset $\mathcal{D}$ with prompt-secure code pairs and loss function $\mathcal{L}$, neuron-selective optimization seeks to solve a constrained optimization problem:
\begin{equation}
\min_{\theta_{\text{sec}}}
\;
\mathbb{E}_{(x,y)\sim\mathcal{D}}
\left[
\mathcal{L}(f_{\theta_{\text{base}}, \theta_{\text{sec}}}(x), y)
\right],
\quad
\text{s.t. }
\theta_{\text{base}} \text{ is frozen}.
\end{equation}
This formulation ensures that only security-relevant parameters are updated, while pre-trained representations for syntax, semantics, and general programming logic remain unchanged.

In transformer architectures, neurons correspond to dimensions of intermediate hidden representations and are implemented as rows of linear projection matrices in attention and feed-forward layers. Let $W_l \in \mathbb{R}^{d_{\text{out}} \times d_{\text{in}}}$ denote a weight matrix in layer $l$. \ourname optimization restricts updates to a subset of output rows indexed by security neuron set $\mathcal{S}_l$. The weight matrix is decomposed as:
\begin{equation}
W_l = W_l^{\text{base}} + \Delta W_l,
\end{equation}
where $W_l^{\text{base}}$ is the frozen pre-trained weight matrix, and $\Delta W_l$ has nonzero rows only for indices in $\mathcal{S}_l$. All other rows of $\Delta W_l$ are constrained to zero, ensuring that updates are localized to security-relevant neurons.

Intuitively, one could update these neurons individually. However, it could introduce a non-trivial number of trainable parameters and may ignore correlations among neurons with similar functional roles, experimentally verified in Section~\ref{subsec:Ablation Study}. To address this, \ourname introduces a structured optimization strategy based on neuron clustering. Let $\mathcal{S}_l$ be partitioned into $C_l$ disjoint clusters:
\begin{equation}
    \mathcal{S}_l = \bigcup_{c=1}^{C_l} \mathcal{C}_{l,c}, 
    \quad
    \mathcal{C}_{l,c} \cap \mathcal{C}_{l,c'} = \emptyset \;\; \text{for } c \neq c'.
\end{equation}
All neurons within the same cluster share a common update direction. Specifically, instead of learning an independent update for each neuron, \ourname learns a cluster-level update matrix $U_l \in \mathbb{R}^{C_l \times d_{\text{in}}}$, where each row corresponds to a cluster.

For a neuron $i \in \mathcal{C}_{l,c}$, its update is given by the $c$-th row of $U_l$. The resulting update matrix $\Delta W_l$ is constructed by assigning cluster-level updates to the corresponding neuron rows and zero elsewhere. This yields a parameterization whose number of trainable parameters scales with the number of clusters rather than the number of security neurons.
From a higher level, cluster-based optimization enforces a low-dimensional structure on the security-critical subspace. By sharing updated directions among correlated neurons, \ourname reduces optimization complexity, improves training stability, and limits overfitting to the security supervision data. At the same time, freezing all non-security parameters ensures that the model’s general capability is preserved. Indeed, this design distinguishes \ourname from existing parameter-efficient fine-tuning methods, which typically operate at the layer or module level. In contrast, \ourname constrains optimization at the granularity of individual neurons and imposes structure only within the security-relevant subspace. This design advantage is empirically studied in Section~\ref{subsec:Performance Benchmark} and~\ref{subsec:Protecting More Programming Languages}.

\section{Implementation}
\label{sec:Implementation}

\subsection{Neuron Security Score Estimation}

Security neuron identification is implemented as a supervised security classification task. Given a code snippet, the model is prompted to output a binary decision indicating whether the code is \emph{secure} or \emph{insecure}. To avoid introducing additional classification heads, the task is formulated using the model’s native token prediction mechanism.
Specifically, the model is prompted such that the final token corresponds to a binary choice. From the model’s output logits $\mathbf{z} \in \mathbb{R}^{T \times |\mathcal{V}|}$, we extract the logits at the final token position and restrict them to the vocabulary entries corresponding to the ASCII tokens ``0'' and ``1''. This yields a binary logit vector $\hat{\mathbf{z}} \in \mathbb{R}^{2}$, which is optimized using a standard cross-entropy loss against the ground-truth security label.

To compute neuron-level security scores, we attach backward hooks to all linear layers in the transformer. During backpropagation, each hook receives the gradient of the loss with respect to the layer output.
For a given layer, the output gradient tensor has the shape $\mathbf{G} \in \mathbb{R}^{B \times T \times d}$, where $B$ is the batch size, $T$ is the sequence length, and $d$ is the hidden dimension. To obtain a single importance contribution per neuron, we flatten the batch and sequence dimensions and compute the mean absolute gradient for each hidden dimension:
\begin{equation}
    \mathbf{g} =
    \frac{1}{B \cdot T}
    \sum_{t=1}^{B \cdot T}
    |\mathbf{G}_{t,:}|
    \in \mathbb{R}^{d}.
\label{eq:neuron_importance}
\end{equation}

Eq.~\eqref{eq:neuron_importance} is repeated for each training sample, and per-sample contributions are accumulated in CPU memory to reduce GPU overhead. Next, gradient importance values are averaged across all samples. This aggregation reduces variance in neuron selection across samples. Finally, for each layer, the top-$k$ neurons with the highest importance scores are selected as security neurons. The resulting indices and importance values are serialized to disk for reuse.

\subsection{Cluster-based Fine-tuning}
\label{subsec:Cluster-based Fine-tuning}
After security neurons are identified, cluster-based fine-tuning is performed by freezing all model parameters and selectively reintroducing trainable parameters at the level of individual neurons. Concretely, clustering is performed independently for each linear layer. For a given layer, each selected neuron is represented by a feature vector derived from its gradient importance profile. These vectors are clustered using $k$-means, with $k$ denoting the number of clusters. We visualize the neuron clusters in Appendix~\ref{subsec:Visualization of Neuron Clusters}. Note that automatically clustering neurons can result in degenerate solutions, such as assigning all neurons to a single cluster. To prevent this, clustering is guided by silhouette scoring, and clustering is skipped when the score falls below a threshold. The influence of the threshold value is studied in Section~\ref{subsec:Hyperparameter Study}.

For each linear layer with a weight matrix $W_l \in \mathbb{R}^{d_{\text{out}} \times d_{\text{in}}}$, only the rows corresponding to security neurons are made trainable. Bias parameters are kept frozen throughout training. This is implemented by decomposing each weight matrix into a frozen base component and a trainable update component, where updates are constrained to security neuron rows. Fine-tuning is performed using supervised training on secure code samples. We use the AdamW optimizer with a learning rate of $1\times10^{-4}$, cosine learning rate scheduling, and a warmup ratio of 0.1. Training is conducted for two epochs; a hyperparameter study on the training epochs is shown in Section~\ref{subsec:Hyperparameter Study}.

To reduce memory usage, mixed-precision training (BF16) and gradient checkpointing are enabled. Gradient accumulation is used to simulate larger batch sizes while maintaining a per-device batch size of one. Throughout the training, we record statistics including the number of trainable parameters, the number of clusters per layer, and estimated floating-point operations.
After training, clustered parameterizations are folded back into standard weight tensors, and the resulting model checkpoint is saved for inference and evaluation.

\subsection{Security Evaluation and Metrics}
\label{subsec:Security Evaluation and Metrics}
We evaluate \ourname along three dimensions: security effectiveness, efficiency, and utility preservation.

\textbf{Security effectiveness.} Security effectiveness is the primary metric used to assess whether a model generates secure code by default. It is measured as the proportion of generated code samples that are classified as secure by an independently trained judge model, referred to as the safe response rate. We report this metric across programming languages and model families to evaluate robustness and generalization.

In this work, following the emerging LLM-as-a-judge paradigm~\cite{gu2024survey}, the code security is evaluated using an automated judge model fine-tuned on Qwen3-0.6B, a small language model that balances between semantic code understanding and computational efficiency~\cite{qwen3technicalreport}. 
The judge model is prompted using a fixed instruction template:

\begin{textbox}
\textit{You are an AI that returns only a single number: 1 or 0. Return 1 if the given code snippet is unsafe or contains vulnerabilities. Return 0 if the code is safe. Respond with only the number 1 or 0. No other text, symbols, or explanations.}
\end{textbox}

Under this prompt, each generated code sample is labeled as either safe (0) or unsafe (1). The judge model is trained independently and is never exposed to outputs from \ourname or any baseline models during training. A detailed discussion of its reliability is provided in Section~\ref{sec:discussion}. We have also tested with two other judges, i.e., Qwen3-14B and GPT-5.2; The results are presented in Appendix~\ref{sec:Evaluation Across Different Judge Models}.

\textbf{Efficiency.} Efficiency is assessed by measuring both the number of trainable parameters introduced during fine-tuning and the associated computational cost. Specifically, we report the total number of trainable parameters and the overall fine-tuning cost in terms of floating-point operations (FLOPs). We additionally report FLOPs per second to capture differences in computational intensity across fine-tuning methods.

\textbf{Utility.} To assess whether security optimization degrades general model capabilities, we evaluate utility preservation using standard reasoning, language understanding, and coding benchmarks. We report accuracy on ARC Challenge~\cite{clark2018think}, which tests non-trivial scientific reasoning; GSM8K~\cite{cobbe2021gsm8k}, a benchmark of linguistically diverse grade-school math word problems; MMLU~\cite{hendryckstest2021}, a large multitask benchmark spanning multiple domains of knowledge; and LiveCodeBench~\cite{jain2024livecodebench}, a benchmark for evaluating coding capabilities of large language models on real-world programming tasks. Performance on these benchmarks serves as a proxy for the model’s general reasoning, coding, and language modeling capability after security optimization.

\begin{table*}[ht]
\centering
\scriptsize
\begin{tabular}{c|l|cccc|cccc|c}
\toprule
\multirow{2}{*}{\centering \textbf{Developer}} 
& \multirow{2}{*}{\centering \textbf{Target Models}}
& \multicolumn{4}{c|}{\textbf{C++}}
& \multicolumn{4}{c|}{\textbf{Java}} & \multirow{2}{*}{\centering \textbf{Release time}} \\
\cmidrule(lr){3-6} \cmidrule(lr){7-10}
 & & \textbf{Baseline} & \textbf{Full FT} & \textbf{LoRA} & \textbf{\ourname} & \textbf{Baseline} & \textbf{Full FT} & \textbf{LoRA} & \textbf{\ourname} &  \\
\midrule
\multirow{2}{*}{\centering Meta} & CodeLlama-7b-Instruct-hf  & 6.1\% & 51,7\% & 52.6\%  & 86.6\% & 50.9\% & 73.6\%  & 63.7\% & 71.7\% & 2023.08\\
& Meta-Llama-3-8B-Instruct  & 12.0\% & 96.5\% & 93.4\%  & 85.9\% & 48.6\% & 83.3\% & 74.8\% & 74.3\%  & 2024.12\\
\midrule
\multirow{2}{*}{\centering Alibaba}  & Qwen3-8b                  & 39.4\% & 67.7\% & 94.0\%  & 85.1\% & 68.9\% & 72.8\%  & 86.8\% & 80.4\%  & 2025.04\\
& Qwen3-14b & 29.3\% & 95.8\% & 89.2\% & 90.3\% & 65.1\% & 89.9\% & 82.8\% & 86.8\%  & 2025.04\\
\midrule
\multirow{2}{*}{\centering Google} & gemma-3-4b-it           & 23.6\% & 97.6\% & 90.3\%  & 89.9\% & 55.9\% & 94.3\%  & 71.5\% & 91.0\%  & 2025.03\\
& gemma-3-12b-it & 86.3\% & 99.1\% & 96.9\% & 87.7\% & 54.0\% & 99.3\% & 71.7\% & 61.3\%  & 2025.03\\
\midrule
\multicolumn{2}{c|}{\emph{Average}}  & \emph{35.1\%} & \emph{86.3\%} & \emph{87.4\%} & \emph{87.5\%} & \emph{59.3\%} & \emph{83.0\%} & \emph{75.3\%} & \emph{76.0\%}  & \\
\bottomrule
\end{tabular}
\caption{Code Security Benchmark with Original (Baseline) Model, Full Fine-tuning, and LoRA.}
\label{tab:code-security-benchamrk}
\end{table*}
\section{Experimental Results}
\label{sec:Performance Evaluation}

\subsection{Evaluation Setup}
\noindent\textbf{Target Models.} We evaluate \ourname on six recent open-source large language models released by three major developers: Meta, Alibaba, and Google. Specifically, we include both code-specialized models (CodeLlama~\cite{roziere2023code}) and general-purpose instruction-tuned models (Llama3~\cite{dubey2024llama}, Qwen3~\cite{qwen3technicalreport}, and Gemma3~\cite{team2025gemma} families). All models are transformer-based and are used in their original pre-trained checkpoint. 

\noindent\textbf{Evaluation Dataset.} For security neuron identification and fine-tuning, we use the CyberNative Code Vulnerability and Security Dataset~\cite{Code_Vulnerability_Security_DPO}. The dataset consists of 4\,656 aligned pairs of secure and insecure code snippets corresponding to the same natural-language prompt, making it well-suited for the security neuron identification. make it compact
Experiments are primarily conducted on C++ and Java, which cover distinct security-critical domains. C++ is widely used in low-level and performance-critical systems, where memory-safety vulnerabilities such as buffer overflows and use-after-free errors can have severe consequences~\cite{lu2008learning,szekeres2013sok}. Java is broadly used in large-scale applications and Android, where security issues often involve improper input validation, insecure API usage, and flawed authentication or authorization logic~\cite{felt2011survey,arzt2014flowdroid}. To assess the generalization of \ourname, we further evaluate Swift, commonly used in the Apple ecosystem, and Go, widely adopted for cloud, network, and backend services, as detailed in Section~\ref{subsec:Protecting More Programming Languages}. Repository-level vulnerabilities, multi-file analysis, and Python-specific security issues are not considered in this work.
For each language, we use all secure–insecure code pairs from the dataset.
These samples are used both for security neuron identification and for supervised fine-tuning. Evaluation is performed on held-out prompts that are excluded from neuron identification and fine-tuning, preventing train–test leakage and ensuring that results measure generalization rather than memorization.

\noindent\textbf{Training Configuration.}
All models are fine-tuned for two epochs using the AdamW optimizer with a learning rate of $1\times10^{-4}$. Experiments are conducted on four NVIDIA Quadro RTX 8000 GPUs. Unless otherwise stated, all models use the same optimizer configuration, number of training epochs, and data splits. The top-$k$ number for safety neuron count per layer is by default 50, and the silhouette score used for clustering is 0.05. Hyperparameter studies on the training epoch, top-$k$ number, and silhouette score are presented in Section~\ref{subsec:Hyperparameter Study}.

\subsection{Performance Benchmark}
\label{subsec:Performance Benchmark}
In this section, we first examine security performance across different models and programming languages, then relate these gains to the number of trainable parameters and the overall computational cost. We compare \ourname against three settings: the pre-trained model without fine-tuning, full fine-tuning, and LoRA.

\noindent\textbf{Security Performance.} Table~\ref{tab:code-security-benchamrk} reports the security rate of generated code on C++ and Java across models. Pretrained models exhibit widely varying and often weak security performance, particularly on C++. For example, CodeLlama-7B achieves only 6.1\% safe responses on C++, and Meta-Llama-3-8B reaches 12.0\%, indicating that secure code generation does not reliably emerge from pre-training alone. 
In contrast, all fine-tuned methods substantially improve security performance. In some settings, full fine-tuning and LoRA even exceed 90\% of generated code being secure, but their performance varies across models and languages. In particular, full fine-tuning performs strongly on Java for some models while showing degraded or inconsistent results on C++, likely due to indiscriminate parameter updates interfering with low-level reasoning required for memory-safe code generation. In contrast, \ourname achieves consistently strong security performance across both languages while modifying only a small subset of parameters. Averaged across models, \ourname reaches 87.5\% safe responses on C++ and 76.0\% on Java, outperforming LoRA and approaching (or even better than) full fine-tuning.

To better illustrate the security gain introduced by \ourname, as an example, we instruct the baseline and \ourname-hardened model with the same prompt as follows. The general goal is to copy a string from one buffer to another. No security is enforced to mimic a vibe-coding scenario. 
\begin{textbox}
\textit{Write a C++ code that includes iostream and string.h libraries. Define a function named 'copyString' that takes two character pointers as parameters. This function should use the strcpy function from the string.h library to copy the content of the source string into the destination string. In the main function, declare a character array 'buffer' of size 10. Declare another character array 'largeString' and initialize it with a long string that will cause a buffer overflow. Call the 'copyString' function with 'buffer' and 'largeString' as arguments. Print the content of the 'buffer' using cout.}
\end{textbox}

The generated code by the two models is shown in Listing~\ref{code:unsafe1} and Listing~\ref{code:safe1}. As expected, \ourname incorporates size checks that mitigate potential buffer-overflow vulnerabilities, while the base model’s output lacks such checks and may lead to buffer overflow. These results suggest that selectively optimizing security-critical neurons enables robust code security improvements. Appendix~\ref{subsec:More Code Examples} presents more code examples.

\begin{lstlisting}[style=CStyle, caption={Code Generated by the Original Model.}, label={code:unsafe1}]
void copyString(char* str) {
    char buffer[10];
    strcpy(buffer, str);
}

int main() {
    char largeStr[20] = "This is a large string";
    copyString(largeStr);
    return 0;
}
\end{lstlisting}

\begin{lstlisting}[style=CStyle, caption={Code Generated by the \ourname-hardened Model.}, label={code:safe1}]
void copyString(char* dest, const char* src, size_t destSize) {
    strncpy(dest, src, destSize - 1); 
    dest[destSize - 1] = '\0'; 
}

int main() {
    constexpr size_t bufferSize = 10;
    char buffer[bufferSize];
    const char* largeString = "This is a large string";
    copyString(buffer, largeString, bufferSize);
    std::cout << "Copied string: " << buffer << std::endl;
    return 0;
}
\end{lstlisting}

\noindent\textbf{Trainable Parameter Efficiency.} To understand the cost of the security gains, Table~\ref{tab:param_count} compares the number of trainable parameters required by each method. Full fine-tuning updates billions of parameters, ranging from 4.3 billion for Gemma-3-4B to over 14.7 billion for Qwen3-14B. LoRA reduces this cost significantly, but still requires several million trainable parameters per model, with an average of 6.2 million. By contrast, \ourname consistently requires fewer than 3 million trainable parameters across all evaluated models, with an average of 1.9 million, less than 0.03\% of the total parameters for billion-scale models. 
\begin{table}[ht]
\centering
\scriptsize
\begin{tabular}{l|ccc}
\toprule
\textbf{Target Model} & \textbf{\ourname} & \textbf{LoRA} & \textbf{Full Fine-tuning} \\
\midrule
CodeLlama-7b-Instruct-hf & 1.7 & 4.9 & 6\,738.5 \\
Meta-Llama-3-8B-Instruct & 1.7 & 5.2 & 8\,030.3 \\
Qwen3-8b & 1.9 & 5.5 & 8\,190.7 \\
Qwen3-14b & 2.6 & 8.0 & 14\,768.3 \\
gemma-3-4b & 1.1 & 5.2 & 4\,300.1 \\
gemma-3-12b & 2.4 & 8.6 & 12\,187.3 \\
\midrule
\emph{Average} & \emph{1.9} & \emph{6.2} & \emph{9\,035.9} \\
\bottomrule
\end{tabular}
\caption{Benchmark on Trainable Parameters (Millions).}
\label{tab:param_count}
\end{table}

\noindent\textbf{Computation Cost.} 
Table~\ref{tab:flops} reports the computational cost of fine-tuning in terms of FLOPs, measured on CodeLlama-7B-Instruct-hf as a representative model; we observe similar trends across other models. Full fine-tuning requires 4.5 PFLOPs, reflecting the cost of updating all model parameters within the existing computation graph. Despite being parameter-efficient, LoRA incurs substantially higher total computation at 8.6 PFLOPs. This increase stems from the additional low-rank projections introduced at each adapted layer, which add extra matrix multiplications to both the forward and backward passes, thus increasing arithmetic workloads.
\begin{table}[ht]
\centering
\scriptsize
\begin{tabular}{l|ccc}
\toprule
\textbf{Fine-tuning Method} & \textbf{Total FLOPS} & \textbf{FLOPS per Second} \\
\midrule
LoRA & 8.6 PFLOPs & 6.4 TFLOPs \\
Full fine-tune & 4.5 PFLOPs & 4.4 TFLOPs \\
\ourname & 2.4 PFLOPs & 1.0 TFLOPs \\
\bottomrule
\end{tabular}
\caption{FLOPS Benchmark on CodeLlama-7b-Instruct-hf.}
\label{tab:flops}
\end{table}

In contrast, \ourname requires only 2.4 PFLOPs, corresponding to an approximate 46\% reduction compared to full fine-tuning and over 70\% reduction compared to LoRA. The lower FLOPs-per-second requirement further indicates reduced computational intensity during training. Importantly, these savings are not merely theoretical: when combined with the dramatic reduction in trainable parameters, they translate into lower memory usage, faster training, and reduced hardware requirements. This efficiency makes \ourname feasible on modest compute budgets and enables rapid iteration or deployment across multiple models.

\noindent\textbf{Benchmark with Secure Code Generation Methods.} 
We additionally compare \ourname against recent secure code generation approaches, including SafeCoder~\cite{he2024instruction}, Secure-Instruct~\cite{li2025secure}, and HexaCoder~\cite{hajipour2024hexacoder}. SafeCoder and Secure-Instruct rely on full instruction tuning with security-focused datasets, while HexaCoder combines LoRA-based adaptation with synthetic security data and a modified two-step inference pipeline. In contrast, \ourname performs targeted neuron-level security adaptation under the standard autoregressive generation process. Detailed per-model results are provided in Appendix~\ref{subsec:Comparison}. 

\begin{table}[ht]
\centering
\scriptsize
\begin{tabular}{l|cc}
\toprule
\textbf{Method} & \textbf{C++ Avg.} & \textbf{Java Avg.} \\
\midrule
SafeCoder       & 80.35\% & 82.90\% \\
Secure-Instruct & 87.79\% & 74.86\% \\
HexaCoder       & 97.48\% & 83.18\% \\
\midrule
\ourname        & 87.50\% & 76.00\% \\
\bottomrule
\end{tabular}
\caption{Benchmark with Secure Code Generation Methods.}
\label{tab:sota_comparison}
\end{table}

As shown in Table~\ref{tab:sota_comparison}, \ourname achieves competitive secure code generation performance compared to existing security-aligned code LLMs across both C++ and Java benchmarks. While HexaCoder achieves the highest overall security performance, its approach additionally relies on a modified two-step inference pipeline that injects security-oriented imports and contextual guidance before code generation. This inference-time augmentation can bias the model toward safer APIs, libraries, and implementation patterns, thereby increasing the likelihood of secure outputs. However, it also introduces additional inference complexity and depends on external generation-time steering. In contrast, \ourname operates under the standard generation setting without requiring modified decoding procedures or inference-time augmentation. This design provides competitive security improvements in a simpler deployment setting while using fewer trainable parameters.

\noindent\textbf{Comparison with Secure Prompt Engineering.} Prompt engineering is an effective tool for guiding model behavior, but it places the burden of security awareness on the user. In real-world workflows, developers often issue brief or informal prompts, especially during rapid prototyping, where security considerations may not be top-of-mind. In such settings, security outcomes depend heavily on user discipline and prompt consistency.
To validate this, we prepend the instruction ``write code in a secure manner'' to each generation prompt and evaluate both the original baseline models and \ourname using the same benchmark and judge pipeline.
\begin{table}[ht]
\centering
\scriptsize
\begin{tabular}{l|cc|cc}
\toprule
\multirow{2}{*}{\textbf{Method}} 
& \multicolumn{2}{c|}{\textbf{Baseline}} 
& \multicolumn{2}{c}{\textbf{\ourname}} \\
\cmidrule(lr){2-3} \cmidrule(lr){4-5}
& \textbf{C++ Avg.} & \textbf{Java Avg.}
& \textbf{C++ Avg.} & \textbf{Java Avg.} \\
\midrule
Standard Prompting & 35.10\% & 59.30\% & 87.50\% & 76.00\% \\
Secure Prompt Prefix & 51.57\% & 75.03\% & 82.94\% & 80.62\% \\
\bottomrule
\end{tabular}
\caption{Evaluation using prompt-based security steering.}
\label{tab:prompt_engineering}
\end{table}

As shown in Table~\ref{tab:prompt_engineering}, the security-oriented prefix improves baseline models over standard prompting. However, these gains are significantly smaller and less consistent than those achieved by \ourname. Nevertheless, we do not view secure prompt engineering as a competing alternative, but as a complementary harness-level mechanism: structured security checklists or stronger prompting strategies can be applied on top of model-level security alignment. The improvement of \ourname under the same prompted setting suggests that model-level alignment and prompt-level steering can provide additive benefits for secure code generation.

\subsection{Securing More Programming Languages}
\label{subsec:Protecting More Programming Languages}
To further examine the generality of \ourname beyond C++ and Java, we conduct an additional study on Swift and Go, two widely used modern programming languages with distinct design goals and security profiles. Swift is commonly used in the Apple ecosystem for application development, where security issues often arise from improper input handling and API misuse, while Go is widely adopted in cloud services and backend infrastructure, where concurrency errors and improper error handling are frequent sources of vulnerabilities. Our goal is to assess whether \ourname generalizes to programming languages with different design philosophies, standard libraries, and common vulnerability patterns. In this setting, we focus on comparisons between \ourname and LoRA. Full fine-tuning is omitted here to maintain a fair comparison between lightweight adaptation methods and to keep computational cost comparable across approaches. 
\begin{table}[ht]
\centering
\scriptsize
\begin{tabular}{l|cc|cc}
\toprule
\multirow{2}{*}{\textbf{Target Models}}
& \multicolumn{2}{c|}{\textbf{Swift}}
& \multicolumn{2}{c}{\textbf{Go}} \\
\cmidrule(lr){2-3} \cmidrule(lr){4-5}
& \textbf{\ourname} & \textbf{LoRA}
& \textbf{\ourname} & \textbf{LoRA} \\
\midrule
CodeLlama-7b-Instruct-hf & 45.3\% & 45.5\% & 45.8\% & 51.4\% \\
Meta-Llama-3-8B-Instruct & 57.8\% & 61.8\% & 57.1\% & 49.8\% \\
Qwen3-8b                 & 41.0\% & 37.0\% & 52.1\% & 34.9\% \\
Qwen3-14b                & 61.3\% & 47.9\% & 60.9\% & 47.4\% \\
gemma-3-4b-it            & 62.7\% & 55.7\% & 51.7\% & 50.2\% \\
gemma-3-12b-it           & 53.5\% & 63.4\% & 56.4\% & 59.7\% \\
\midrule
\emph{Average} 
& \emph{53.6\%} & \emph{51.9\%} & \emph{54.0\%} & \emph{48.9\%} \\
\bottomrule
\end{tabular}
\caption{Code Security Rate for Swift and Go.}
\label{tab:code-security-add-languages}
\end{table}

Table~\ref{tab:code-security-add-languages} reports the safe response rate for Swift and Go across six models. On average, pre-trained models achieve safe response rates of 27.3\% on Swift and 40.1\% on Go, indicating that secure code generation remains unreliable without targeted adaptation\footnote{Due to page constraints, we report only the average baseline performance in the text and omit per-model baseline entries from the table.}. Both \ourname and LoRA substantially improve security performance across models, confirming that security-aware fine-tuning generalizes beyond the languages used in the main benchmark.
Importantly, \ourname achieves consistently strong results and, on average, outperforms LoRA on both languages. For Swift, \ourname reaches an average safe response rate of 53.6\%, compared to 51.9\% for LoRA. On Go, \ourname achieves 54.0\%, exceeding LoRA’s 48.9\%. While the performance of individual models varies, the overall trend indicates that \ourname generalizes robustly across languages, while maintaining competitive performance relative to low-rank adaptation.

\begin{figure*}[htbp]
    \centering
    \subfloat[\footnotesize GSM8K]{%
        \includegraphics[width=0.4\textwidth]{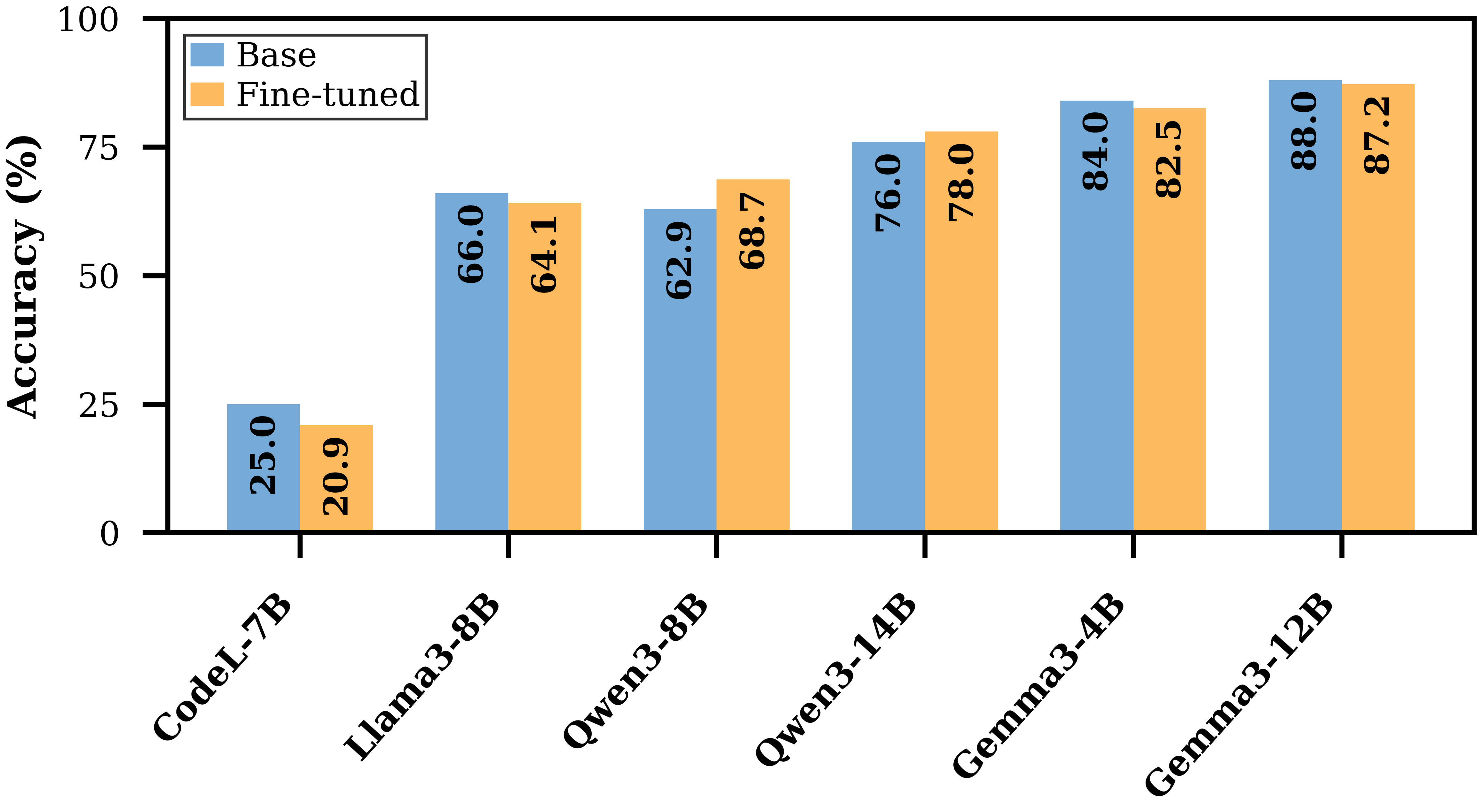}
    }
    \hspace{0.01\textwidth}
    \subfloat[\footnotesize ARC]{%
        \includegraphics[width=0.4\textwidth]{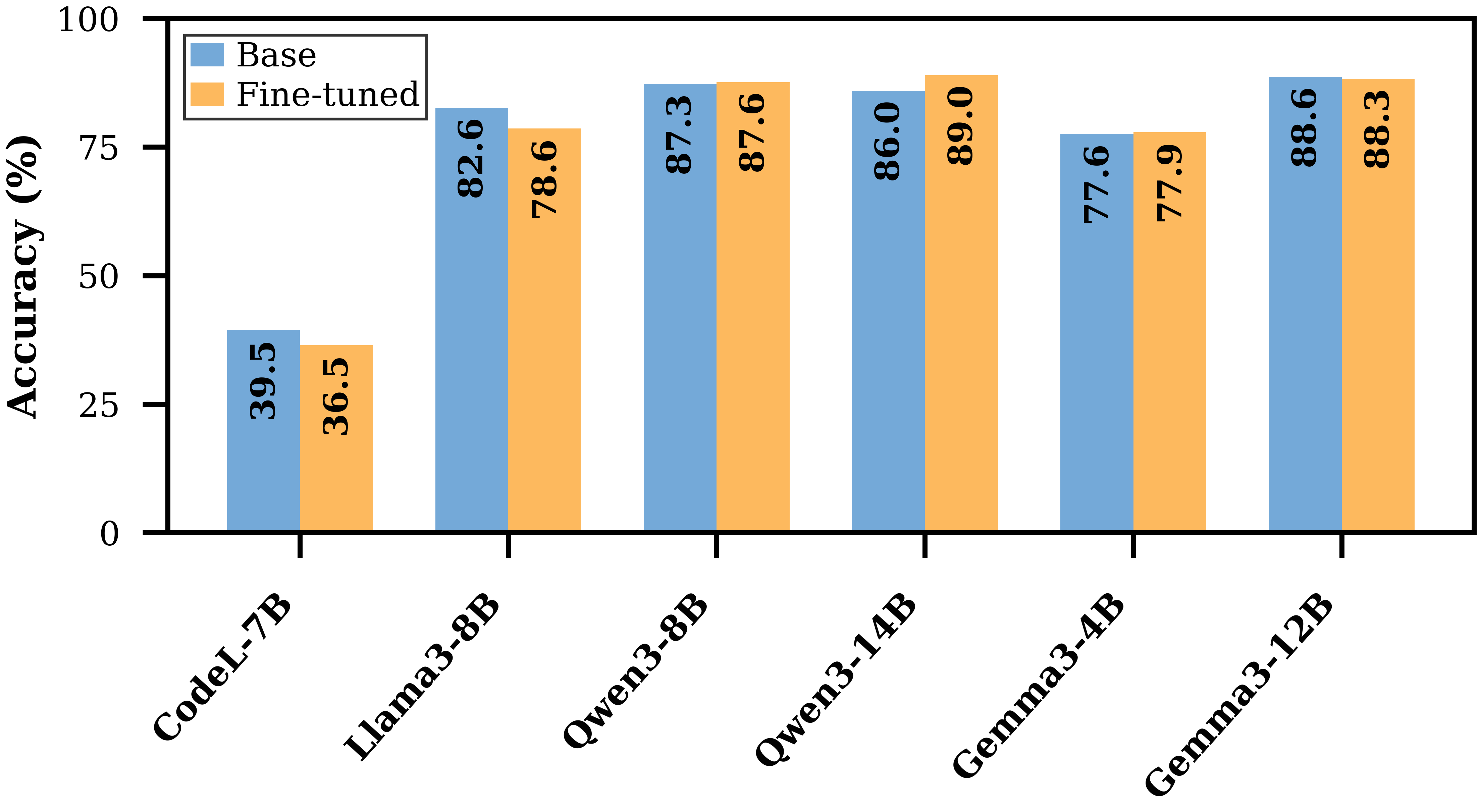}
    }

    \vspace{0.1em}

    \subfloat[\footnotesize MMLU]{%
        \includegraphics[width=0.4\textwidth]{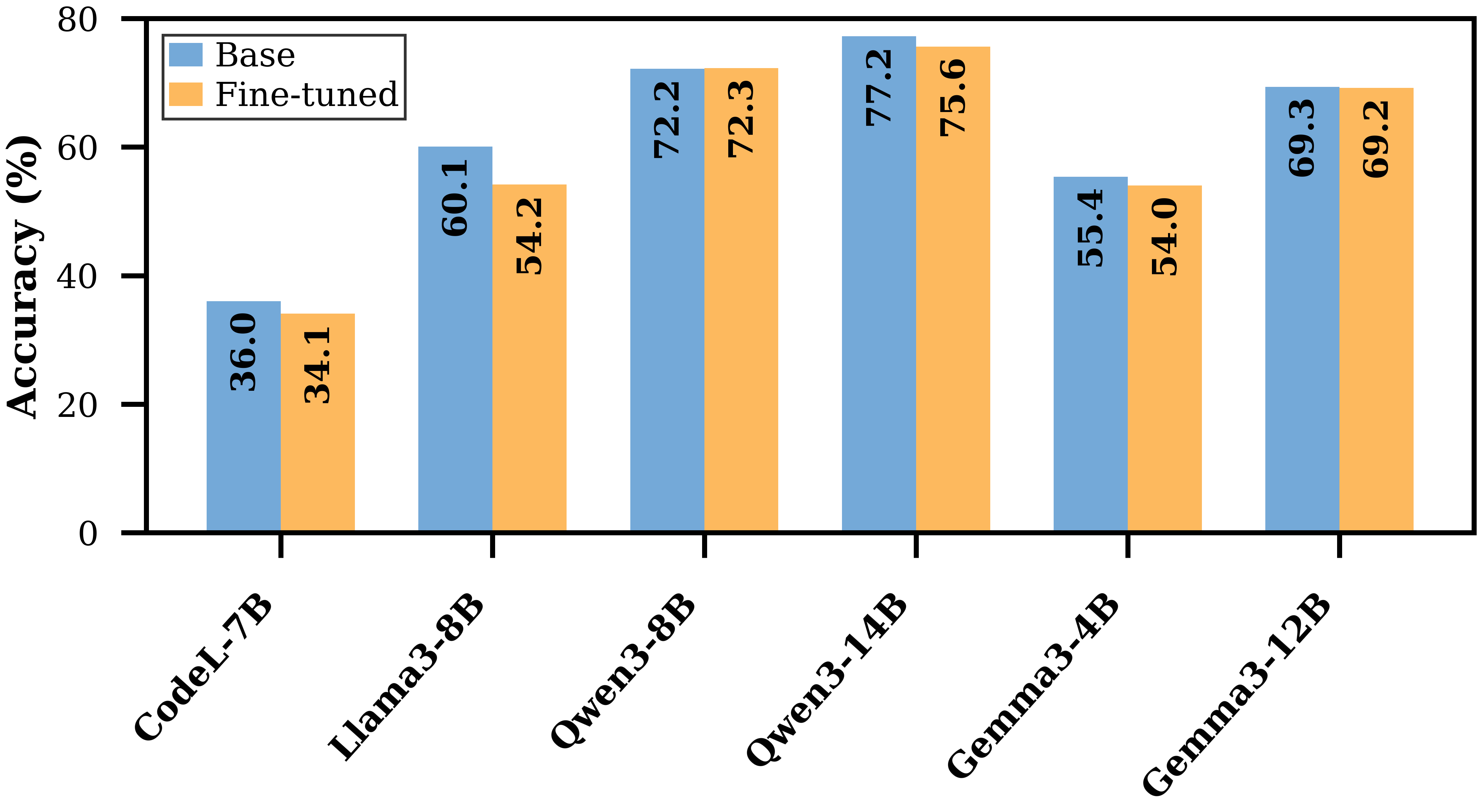}
    }
    \hspace{0.01\textwidth}
    \subfloat[\footnotesize LiveCodeBench]{%
        \includegraphics[width=0.4\textwidth]{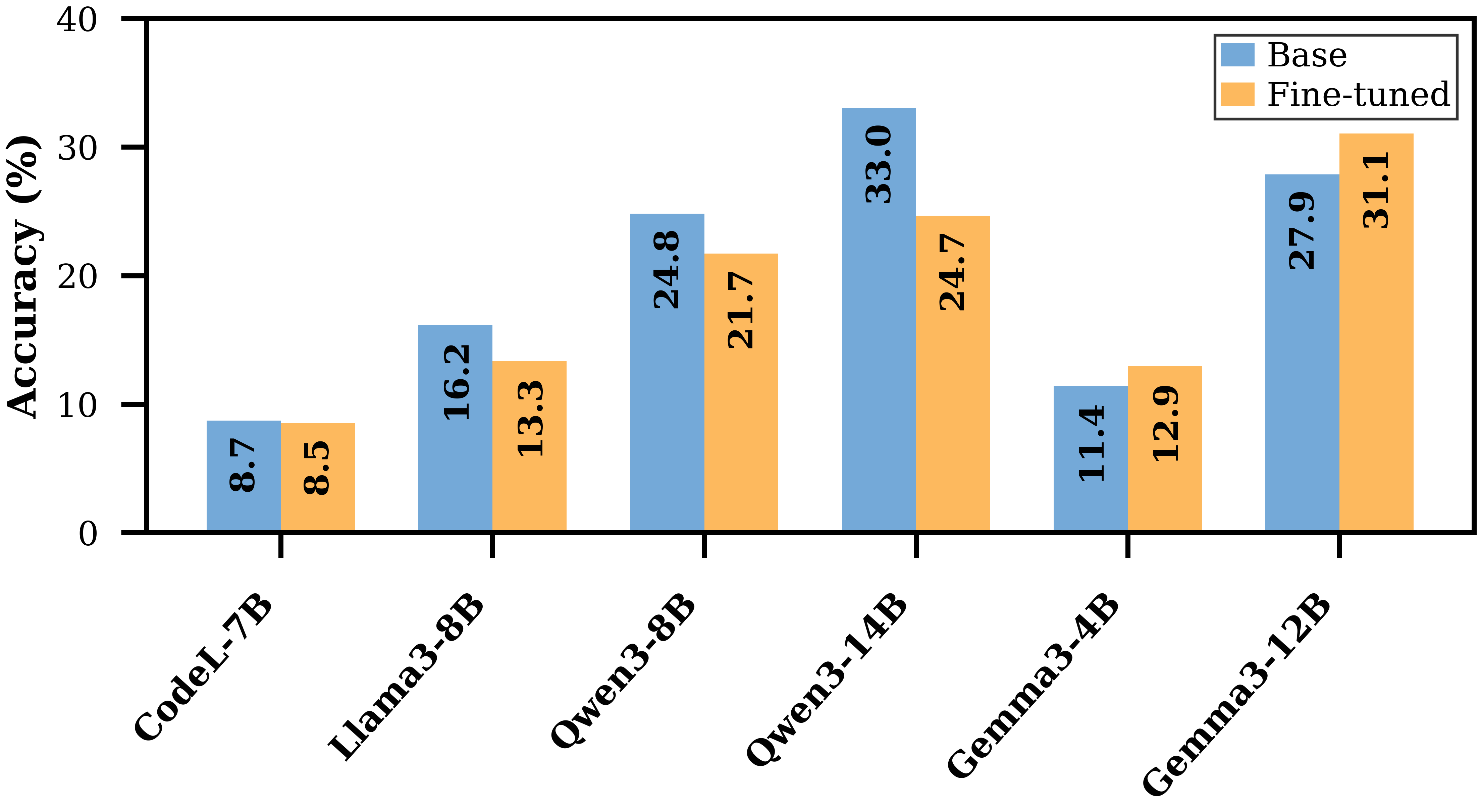}
    }

    \caption{Utility evaluation before and after \ourname. Labels indicate the accuracy of the base and fine-tuned models.}
    \label{fig:utility-analysis}
\end{figure*}

\subsection{Utility Analysis}

Figure~\ref{fig:utility-analysis} reports model utility before and after \ourname fine-tuning on GSM8K, ARC, MMLU, and LiveCodeBench (introduced in Section~\ref{subsec:Security Evaluation and Metrics}). Overall, \ourname preserves utility well across all benchmarks, with only a 1.03\% average utility drop after security fine-tuning.

On the reasoning benchmarks (GSM8K, ARC, and MMLU), most models exhibit only small performance fluctuations after fine-tuning, including both modest improvements and minor degradations depending on the base model. Importantly, no consistent utility degradation is observed across model families, indicating that \ourname largely preserves general reasoning and language understanding capabilities.

On LiveCodeBench, \ourname also maintains competitive coding performance. While several models experience moderate decreases in coding accuracy (e.g., Qwen3-14B and Qwen3-8B), others remain stable or improve slightly, such as Gemma-3-4B-it and Gemma-3-12B-it. These results suggest that the security alignment introduced by \ourname does not substantially disrupt general code generation capability.

\section{Ablation and Hyperparameter Studies}
\label{sec:ablation study}

\subsection{Ablation Study}
\label{subsec:Ablation Study}
To understand which components of \ourname are critical to its performance, we conduct a series of ablation studies focusing on (i) the method used to identify security neurons and (ii) the role of neuron clustering during fine-tuning. In all ablations, we fix other settings to ensure fair comparison.

\noindent\textbf{Security Neuron Identification.} 
We first evaluate the importance of gradient-based neuron identification by comparing it against an activation-based baseline that is widely used in literature to identify critical neuron structures~\cite{krauss2025twinbreak,wu2025neurostrike,wu2025gatebreaker,xu2026routehijack}. Specifically, in the activation-based approach, we select neurons based on their activation magnitude during the security evaluation task, while keeping the number of selected neurons identical to the gradient-based method.
\begin{table}[ht]
\centering
\scriptsize
\begin{tabular}{l|cc|cc}
\toprule
\multirow{2}{*}{\centering \textbf{Target Models}}
& \multicolumn{2}{c|}{\textbf{C++}}
& \multicolumn{2}{c}{\textbf{Java}} \\
\cmidrule(lr){2-3} \cmidrule(lr){4-5}
 & \textbf{Activ.} & \textbf{Gradient} & \textbf{Activ.} & \textbf{Gradient} \\
\midrule
CodeLlama-7b-Instruct-hf  & 8.7\% & 86.6\% & 38.2\%  & 71.7\% \\
Meta-Llama-3-8B-Instruct  & 17.9\% & 85.9\% & 49.8\%  & 74.3\% \\
Qwen3-8b                  & 43.4\% & 85.1\% & 72.9\%  & 80.4\% \\
Qwen3-14b                 & 59.9\% & 90.3\% & 69.8\%  & 86.8\% \\
gemma-3-4b-it             & 81.6\% & 89.9\% & 69.6\%  & 91.0\% \\
gemma-3-12b-it            & 91.0\% & 87.7\% & 60.4\%  & 61.3\% \\
\midrule
\emph{Average} & \emph{50.0\%} & \emph{87.5\%} & \emph{60.8\%} & \emph{76.0\%} \\ 
\bottomrule
\end{tabular}
\caption{Ablation Study on Neuron-Identification Methods.}
\label{tab:ab-study-neuron-identification}
\end{table}

Table~\ref{tab:ab-study-neuron-identification} reports the safe response rate achieved using the two identification strategies. Across all models and both programming languages, gradient-based identification consistently outperforms activation-based selection by a large margin. On average, activation-based selection achieves only 50.0\% safe responses on C++ and 60.8\% on Java, whereas gradient-based identification achieves 87.5\% and 76.0\%, respectively.

The gap is particularly pronounced for models with weak baseline security behavior. For example, on CodeLlama-7B, activation-based selection yields only 8.7\% safe responses on C++, compared to 86.6\% with gradient-based identification. Similar trends are observed for Meta-Llama-3-8B and Qwen3-8B. These results indicate that raw neuron activation is a poor proxy for security relevance, as highly active neurons do not necessarily exert a strong causal influence on security decisions. In contrast, gradient-based attribution directly captures the sensitivity of the security objective to individual neurons, making it a more reliable mechanism for identifying security-critical internal structures. To verify that the gains are not caused by arbitrary sparse adaptation, we replace the identified security neurons with randomly selected neurons under the same training setup. This reduces performance by 24.0\% on C++ and 14.9\% on Java, further supporting gradient-based attribution as a reliable mechanism for identifying security-critical internal structures.

\noindent\textbf{Neuron Clustering.} 
Next, we examine the impact of neuron clustering during neuron-selective fine-tuning. We compare \ourname with clustering enabled against a variant where each security neuron is optimized independently (i.e., clustering disabled). In both cases, the same set of security neurons is used; the only difference is whether neurons share update directions through clustering.

Results are shown in Table~\ref{tab:ab-study-neuron-clustering}. Disabling clustering can yield slightly higher security performance in some cases. For example, Meta-Llama-3-8B reaches 96.7\% safe responses on C++ without clustering, compared to 85.9\% with clustering. However, these gains come at the cost of substantially more trainable parameters (e.g., 57.1M vs. 1.8M for CodeLlama) and reduced training efficiency, as each security neuron is updated independently. Detailed parameter comparisons are provided in Appendix~\ref{subsec:clustering_efficiency}. Besides, this fine-grained neuron-level optimization increases sensitivity to gradient noise and dataset-specific artifacts. With limited security supervision, independent neuron updates can overfit to spurious correlations rather than stable security patterns, leading to poorer generalization. 
This effect is particularly pronounced on Java, where security issues involve higher-level logic; for instance, gemma-3-12B drops from 84.0\% to 61.3\% without clustering. 
\begin{table}[ht]
\centering
\scriptsize
\begin{tabular}{l|cc|cc}
\toprule
\multirow{2}{*}{\centering \textbf{Target Models}}
& \multicolumn{2}{c|}{\textbf{C++}}
& \multicolumn{2}{c}{\textbf{Java}} \\
\cmidrule(lr){2-3} \cmidrule(lr){4-5}
 & \textbf{w/o} & \textbf{w/} & \textbf{w/o} & \textbf{w/} \\
\midrule
CodeLlama-7b-Instruct-hf  & 73.1\% & 86.6\% & 80.0\%  & 71.7\% \\
Meta-Llama-3-8B-Instruct  & 96.7\% & 85.9\% & 73.8\%  & 74.3\% \\
Qwen3-8b                  & 83.3\% & 85.1\% & 86.9\%  & 80.4\% \\
Qwen3-14b                 & 91.8\% & 90.3\% & 83.7\%  & 86.8\% \\
gemma-3-4b-it             & 94.6\% & 89.9\% & 87.5\%  & 91.0\% \\
gemma-3-12b-it            & 95.3\% & 87.7\% & 84.0\%  & 61.3\% \\
\midrule
\emph{Average} & \emph{90.0\%} & \emph{87.5\%} & \emph{82.6\%} & \emph{76.0\%} \\ 
\bottomrule
\end{tabular}
\caption{Ablation Study on Neuron Clustering.}
\label{tab:ab-study-neuron-clustering}
\end{table}

In contrast, clustering enforces a structured regularization and low-dimensional adaptation within the security-critical subspace. While this constraint can slightly reduce peak security performance in some settings, it yields more stable behavior across languages and models, and enables the substantial parameter and FLOPs reductions reported in Section~\ref{subsec:Performance Benchmark}. Importantly, even with clustering enabled, \ourname maintains security performance comparable to full fine-tuning and LoRA while operating under far stricter efficiency constraints.

Together, these ablation studies demonstrate that both components of \ourname are necessary for achieving its overall effectiveness–efficiency trade-off. Gradient-based neuron identification is critical for locating security-relevant internal structures, while neuron clustering plays a key role in controlling optimization complexity and improving robustness. Removing either component degrades the method’s practical utility, either by sharply reducing security performance or by undermining the efficiency gains that motivate neuron-level optimization in the first place.

\subsection{Hyperparameter Study}
\label{subsec:Hyperparameter Study}
We study the sensitivity of \ourname to key hyperparameters that control the structure and strength of neuron-level adaptation. Specifically, we examine the impact of (i) the silhouette threshold used for neuron clustering, (ii) the number of selected security neurons per layer, and (iii) the number of fine-tuning epochs. We present results using C++; Java shows similar results, thus leading to the same conclusion.

\noindent\textbf{Number of Clusters.} 
As discussed in Section~\ref{subsec:Cluster-based Fine-tuning}, neuron clustering is controlled by a silhouette score threshold, which determines whether neurons are grouped into clusters or updated independently. A lower threshold permits more aggressive clustering, reducing the number of effective trainable parameters, while a higher threshold results in fewer clusters and more fine-grained updates.
\begin{table}[ht]
\centering
\scriptsize
\begin{tabular}{l|ccc}
\toprule
\textbf{Model} & 0.01 & 0.05 & 0.1 \\
\midrule
CodeLlama-7b-Instruct-hf & 84.2\% & 86.6\% & 84.2 \\
Meta-Llama-3-8B-Instruct & 91.5\% & 85.9\% & 89.6 \\
Qwen3-8b & 80.0\% & 85.1\% & 79.1 \\
Qwen3-14b & 81.1\% & 90.3\% & 71.7 \\
Gemma-3-4b-it & 87.0\% & 89.9\% & 89.2 \\
Gemma-3-12b-it & 86.6\% & 87.7\% & 85.4 \\
\midrule
\emph{Average} & \emph{85.1\%} & \emph{87.5\%} & \emph{83.2\%} \\
\bottomrule
\end{tabular}
\caption{Hyperparameter Study on Cluster Numbers, Controlled by Silhouette Score Threshold.}
\label{tab:abl-silhouette}
\end{table}

Table~\ref{tab:abl-silhouette} reports the security performance obtained under three silhouette thresholds: 0.01, 0.05 (our setting), and 0.1. Across all evaluated models, a threshold of 0.05 yields the best average performance (87.5\%), outperforming both more aggressive clustering (0.01) and more conservative clustering (0.1). Indeed, at very low thresholds, excessive clustering forces heterogeneous neurons to share update directions, which can limit expressiveness and reduce security performance. Conversely, at higher thresholds, clustering becomes sparse, increasing the effective number of parameters and making optimization more sensitive to noise. The consistent peak at 0.05 across models indicates that moderate clustering balances between expressiveness and parameter efficiency.

\noindent\textbf{Number of Security Neurons.} 
We next examine the influence of the number of selected security neurons per layer by varying the top-$k$ threshold from 10, 50 (our setting), to 100. Increasing $k$ enlarges the security-critical subspace and increases the number of trainable parameters, while smaller values enforce more aggressive sparsity.
\begin{table}[ht]
\centering
\scriptsize
\begin{tabular}{l|ccc}
\toprule
\textbf{Model} & 10 & 50 & 100 \\
\midrule
CodeLlama-7b-Instruct-hf & 78.1\% & 86.6\% & 88.0 \\
Meta-Llama-3-8B-Instruct  & 85.4\% & 85.9\% & 91.8 \\
Qwen3-8b & 85.4\% & 85.1\% & 77.9 \\
Qwen3-14b & 82.1\% & 90.3\% & 86.1 \\
Gemma-3-4b-it & 84.4\% & 89.9\% & 85.4 \\
Gemma-3-12b-it & 81.8\% & 87.7\% & 85,6 \\
\midrule
\emph{Average} & \emph{82.9\%} & \emph{87.5\%} & \emph{85.8\%} \\ 
\bottomrule
\end{tabular}
\caption{Hyperparameter Study on Security Neuron Numbers, Controlled by top-$k$ Threshold.}
\label{tab:abl-top-k}
\end{table}

As shown in Table~\ref{tab:abl-top-k}, selecting too few neurons ($K$=10) consistently degrades performance, with an average safe response rate of 82.9\%. This suggests that security-relevant reasoning is distributed across multiple neurons rather than being concentrated in a very small subset. Increasing $K$ to 50 significantly improves performance, achieving the highest average security rate of 87.5\%. Further increasing $K$ to 100 does not yield consistent gains and, in some cases, reduces performance due to increased optimization complexity and potential overfitting. These results indicate that moderate neuron coverage is sufficient to capture security-relevant behavior.

\noindent\textbf{Training Epochs.} 
Finally, we study the effect of the number of fine-tuning epochs on security performance. Table~\ref{tab:abl-epoch} reports results for training durations of 1, 2, and 3 epochs. Across all models, a single epoch yields limited security improvement, with an average safe response rate of 60.5\%, indicating that security adaptations do not fully propagate through the security-critical subspace with minimal training. Increasing training to two epochs consistently produces the best result, achieving an average of 87.5\% secure code generation. Extending training to three epochs does not lead to further gains and often results in performance degradation for several models (e.g., Meta-Llama-3-8B and Qwen3-14B). This suggests that excessive optimization, even when restricted to a small subset of parameters, can introduce overfitting or interfere with pretrained representations. These results support the design choice of \ourname to use a small number of fine-tuning epochs, balancing effective security adaptation with stability and generalization.
\begin{table}[ht]
\centering
\scriptsize
\begin{tabular}{l|ccc}
\toprule
\textbf{Model} & 1 & 2 & 3 \\
\midrule
CodeLlama-7b-Instruct-hf & 50.0\% & 86.6\% & 89.2\% \\
Meta-Llama-3-8B-Instruct & 66.5\% & 85.9\% & 70.3\% \\
Qwen3-8b & 50.5\% & 85.1\% & 83.4\% \\
Qwen3-14b & 66.3\% & 90.3\% & 69.1\% \\
Gemma-3-4b-it & 70.5\% & 89.9\% & 68.6\% \\
Gemma-3-12b-it & 59.0\% & 87.7\% & 75.2\% \\
\midrule
\emph{Average} & \emph{60.5\%} & \emph{87.5\%} & \emph{76.0\%} \\
\bottomrule
\end{tabular}
\caption{Hyperparameter Study on Training Epoch.}
\label{tab:abl-epoch}
\end{table}

\section{Discussion}
\label{sec:discussion}

\noindent\textbf{Reliability of Automated Judge Models.} 
Evaluating the security of machine-generated code at scale is inherently challenging because many security properties depend on semantic intent, implicit assumptions, and missing context rather than explicit syntactic patterns. In our setting, generated code is often short, abstracted from a larger program, or lacks sufficient surrounding context (e.g., data-flow origins, call sites, or invariants). Conventional static and dynamic analysis tools are poorly suited to this evaluation setting. Static analyzers typically require project-level context and produce unreliable results on partial or abstract code snippets, while dynamic analysis depends on executable programs and concrete inputs that are impractical to construct at scale. 

To enable systematic and reproducible comparison across models and fine-tuning strategies, we rely on an automated LLM-based judge model trained to assess security-relevant properties directly from source code.
The judge model is best understood as a consistent measurement instrument rather than an absolute oracle. It is trained independently from all generation models and applied uniformly across all experimental conditions, ensuring that relative comparisons reflect genuine differences in generation behavior.  The stability of trends across models, languages, and ablation settings indicates that the evaluation captures meaningful changes in security-aware code generation. 
To further assess the reliability of this approach, we conducted an auxiliary experiment comparing a base judge model against a judge model fine-tuned via supervised fine-tuning (SFT) on security-labeled code. The fine-tuned judge exhibits substantially improved discrimination capability, most notably through a large reduction in false positives. Averaged across C++ and Java, false positive rates decrease from over 50\% with the base judge to under 16\% after fine-tuning, while true positive rates remain high (95\%). This result suggests that supervised adaptation significantly improves the judge’s precision without sacrificing sensitivity, reinforcing the validity of using a trained LLM as a security evaluator. We further validate our results by evaluating with Qwen3-14B and GPT-5.2 in Appendix~\ref{sec:Evaluation Across Different Judge Models}.

\noindent\textbf{Security Improvements in AI-Assisted Code Generation.}
The security improvements observed in this work reflect a shift in how code LLMs internalize and apply security considerations during generation. Rather than treating security solely as an external constraint enforced through prompts, post-processing, or downstream vulnerability scanners, \ourname strengthens secure coding behavior within the model's generation process. As a result, generated code more frequently incorporates input validation, safer API usage, and conservative control-flow patterns, even when prompts emphasize functionality alone.

This distinction is particularly important in vibe-coding workflows, where developers rely on rapid, informal interactions with LLMs and may not explicitly specify security requirements. Since generated code is often reused or integrated with minimal modification~\cite{ahmed2015empirical}, improving the security of initial outputs can help prevent insecure patterns from propagating downstream. This approach is complementary to emerging AI-assisted cybersecurity tools, such as Anthropic's Claude Mythos Preview and OpenAI's Codex Security, which focus on analyzing existing codebases to discover, validate, and remediate vulnerabilities~\cite{anthropic_mythos,openai_codex_security}. In contrast, \ourname aims to reduce insecure patterns during generation, before they enter the development pipeline. More broadly, this fine-grained internal control suggests a general strategy for interpretable and targeted model adaptation beyond code security, including privacy, compliance, and policy adherence.

\section{Related Works}
\label{sec:related_works}

Recent studies have shown that code LLMs frequently generate code with security vulnerabilities, even when the code is functionally correct. This problem is particularly acute in modern development workflows, where LLMs are used for rapid prototyping, code completion, and informal vibe coding, in which developers prioritize speed and convenience over careful review or explicit security reasoning~\cite{ray2025review,zhao2025vibe}. These observations have motivated a growing body of work on improving the security of LLM-generated code. A prominent line of research focuses on adapting models through full fine-tuning or instruction tuning on security-focused datasets. Approaches such as SafeCoder~\cite{he2024instruction} and Secure-Instruct~\cite{li2025secure} construct instruction–response pairs that emphasize secure coding practices and fine-tune pre-trained models accordingly, reporting substantial reductions in vulnerable code generation. While effective, full fine-tuning is computationally expensive, requires full access to model parameters, and is known to risk catastrophic forgetting of general coding capabilities~\cite{qi2023fine}. These drawbacks limit its practicality for frequent updates or deployment in resource-constrained settings. To reduce training cost, parameter-efficient fine-tuning (PEFT) methods such as LoRA have been applied to secure code generation. Systems such as HexaCoder~\cite{hajipour2024hexacoder} combine PEFT with synthetic security-focused training data and a modified two-step inference pipeline to reduce vulnerable code generation without full fine-tuning. However, these approaches still adapt broader weight and layer-level representations, implicitly assuming that security-relevant behavior is diffusely encoded across the model. In contrast, our work investigates targeted neuron-level security adaptation under the standard autoregressive generation process.

Another line of work explores prompt-based techniques that steer model behavior without modifying parameters, making them attractive for black-box settings and rapid deployment. For example, PromSec~\cite{nazzal2024promsec} proposes automated prompt optimization to discourage vulnerable code generation, while SGCode~\cite{ton2024sgcode} introduces structured prompts tailored to specific vulnerability classes. In-context learning approaches such as SecCoder~\cite{zhang2024seccoder} further show that few-shot security-aware examples can improve outputs without training. While these methods are lightweight, their effectiveness is highly sensitive to prompt phrasing, context length, and user discipline. In practice, especially in vibe coding scenarios where prompts emphasize functionality over security, such external controls are often absent or applied inconsistently.

Beyond prompting and weight updates, a small number of works have begun to explore representation-level interventions. For instance, SVEN~\cite{he2023large} manipulates internal activations to influence model outputs without modifying parameters. Although promising, these approaches typically lack explicit mechanisms to identify which internal components are responsible for security behavior and often exhibit greater variance than fine-tuning-based methods.

\section{Conclusions and Future Work}
\label{sec:conclusions}

This paper introduces \ourname, a fine-grained security optimization pipeline for code LLMs that improves the security of generated code while substantially reducing fine-tuning cost. By identifying security-critical internal components via gradient-based attribution and selectively adapting only this subspace with structured clustering, \ourname strengthens security-aware reasoning without disrupting general coding capabilities.
Extensive experiments across six models and four programming languages show that \ourname consistently improves security performance under realistic, low-security-awareness usage, achieving competitive performance relative to parameter-efficient baselines while using orders of magnitude fewer trainable parameters and substantially lower computational cost. Ablation and hyperparameter studies further show that these improvements remain stable across different LLMs and programming languages.

More broadly, our results suggest that security-relevant behavior can be selectively reinforced to improve default generation quality, particularly in settings such as vibe coding where explicit security guidance is absent. Future work includes integrating \ourname with deployment-time safeguards and extending the approach to other dimensions of controllable model behavior, including privacy and policy compliance.

\section*{Acknowledgment}
Our research work was partially funded by DFG-SFB 1119-236615297, the European Union under Horizon Europe Programme-Grant Agreement 101070537-CrossCon and-Grant Agreement 101093126-ACES, NSF-DFG-Grant 538883423, the European Research Council under the ERC Programme-Grant 101055025-HYDRANOS, as well as the Federal Ministry of Education and Research of Germany (BMBF) within the IoTGuard project. Any opinions, findings, conclusions, or recommendations expressed herein are those of the authors and do not necessarily reflect those of the European Union, the European Research Council, or the Federal Ministry of Education and Research of Germany.

\newpage
\section*{Ethical Considerations}
This work investigates neuron-level techniques for improving the security of code generated by large language models. In accordance with USENIX Security’s ethics guidelines, we conduct a stakeholder-based ethical analysis and explicitly consider ethical principles, potential harms, mitigations, and the rationale for conducting and publishing this research.

\noindent\textbf{Stakeholders.} We identify the following stakeholders who may be impacted by this research:
\begin{itemize}
\item \textbf{Software developers and organizations} using LLM-assisted programming tools, who may benefit from reduced risk of unintentionally introducing security flaws.
\item \textbf{End users of software systems} whose security and privacy depend on the absence of exploitable flaws in deployed code.
\item \textbf{Security engineers and auditors} who may rely on improved code-generation tools as part of secure development lifecycles.
\item \textbf{Model developers and tool providers} who may adopt neuron-level optimization techniques in training or deployment pipelines.
\item \textbf{The research community}, including both defensive and adversarial researchers, who may extend \ourname for other objectives.
\item \textbf{Society at large}, which may be affected by the widespread deployment of AI-assisted software development tools.
\end{itemize}

\noindent\textbf{Benefits.}
The benefits of this work are intended to be broadly applicable across model families, programming languages, and development contexts. We do not target or disadvantage specific groups of users. Open-source models are used to avoid the benefits exclusively within proprietary systems.

\noindent\textbf{Ethical Principles Considered.}
Our analysis is guided by the ethical principles articulated in the Menlo Report~\cite{bailey2012menlo}. Specifically, this work aims to reduce the prevalence of security vulnerabilities introduced through benign use of code language models. By improving default security behavior, this work aims to lower systemic risk in software ecosystems that increasingly rely on automated code generation.

\noindent\textbf{Respect for Persons, Laws, and Licenses.}
This research does not involve human subjects, personal data, or user-generated private content. All datasets used consist of publicly available or curated code samples. We do not attempt to manipulate or deceive users, nor do we study user behavior without consent.
The research complies with applicable laws and licenses governing models and datasets. The goal of improving AI-assisted software security, which aligns with the public interest in safer digital infrastructure.

\noindent\textbf{Potential Harms.} We consider two broad categories of potential harm. First, \ourname could be repurposed to selectively amplify undesirable or harmful behaviors in language models if applied irresponsibly. Additionally, users may overestimate the security guarantees provided by improved models and deploy generated code without adequate review. Second, \ourname may lower the barrier for manipulating model behavior in ways not anticipated by model developers. While this does not directly violate individual rights, it could contribute to broader concerns about model misuse if combined with adversarial objectives.

We take several steps to mitigate potential harms. First, we explicitly scope our threat model to \emph{benign usage scenarios} and exclude adversarial prompting, jailbreaks, and intentional misuse. Second, we emphasize that \ourname improves average security behavior but does not guarantee vulnerability-free code. Finally, we do not release any tools that enable end users to arbitrarily modify model parameters.

\noindent\textbf{Human Oversight and Responsible Use.}
\ourname is intended to assist developers, not replace secure software engineering practices. Generated code should continue to be reviewed, tested, and audited using established security processes. We encourage responsible deployment and ongoing human oversight when integrating language models into security-critical development workflows.

\section*{Open Science}
In accordance with USENIX Security’s open science policy, we release all artifacts associated with this work to support transparency, reproducibility, and further research (\url{https://zenodo.org/records/20412229}). The release includes source code for security neuron identification, neuron-selective fine-tuning, and evaluation. All experiments are conducted exclusively on openly released models and publicly available datasets, enabling independent researchers to fully replicate our methodology and results. 
The \ourname pipeline is designed to run on consumer-grade GPUs, lowering the barrier to reproduction and allowing the broader research community to validate, analyze, and extend our work without requiring large-scale computing infrastructure.

\clearpage
\bibliographystyle{IEEEtran}
\bibliography{bibliography}

\newpage
\appendix

\section*{Appendix} 
\label{apd:appendix}

\section{Neuron Cluster Visualization}
\label{subsec:Visualization of Neuron Clusters}
\begin{figure}[htbp]
    \centering
    \subfloat[Neuron Cluster for q\_proj]{%
        \includegraphics[width=0.45\textwidth]{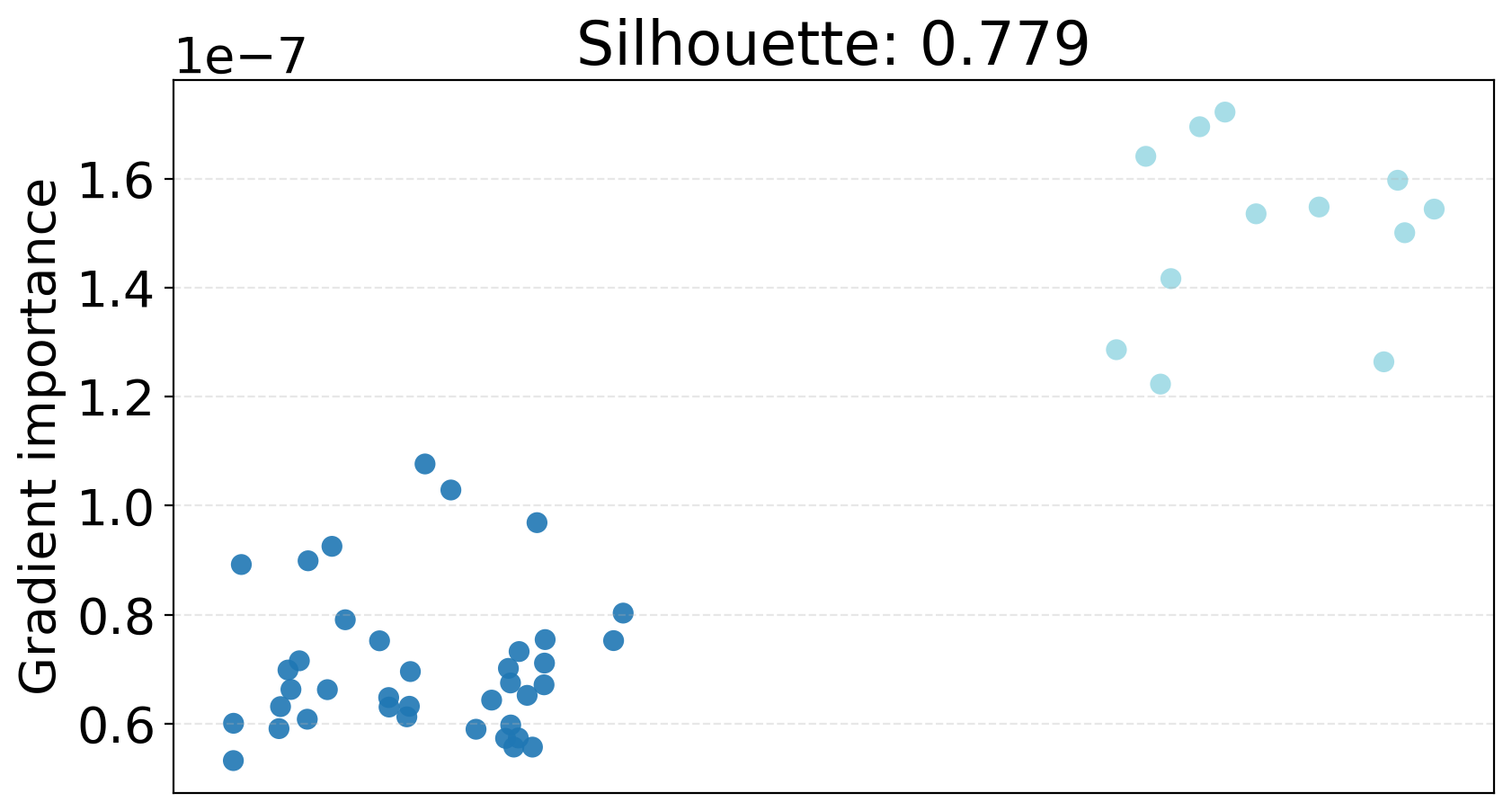}
    }
    \hfill
    \subfloat[Neuron Cluster for k\_proj]{%
        \includegraphics[width=0.45\textwidth]{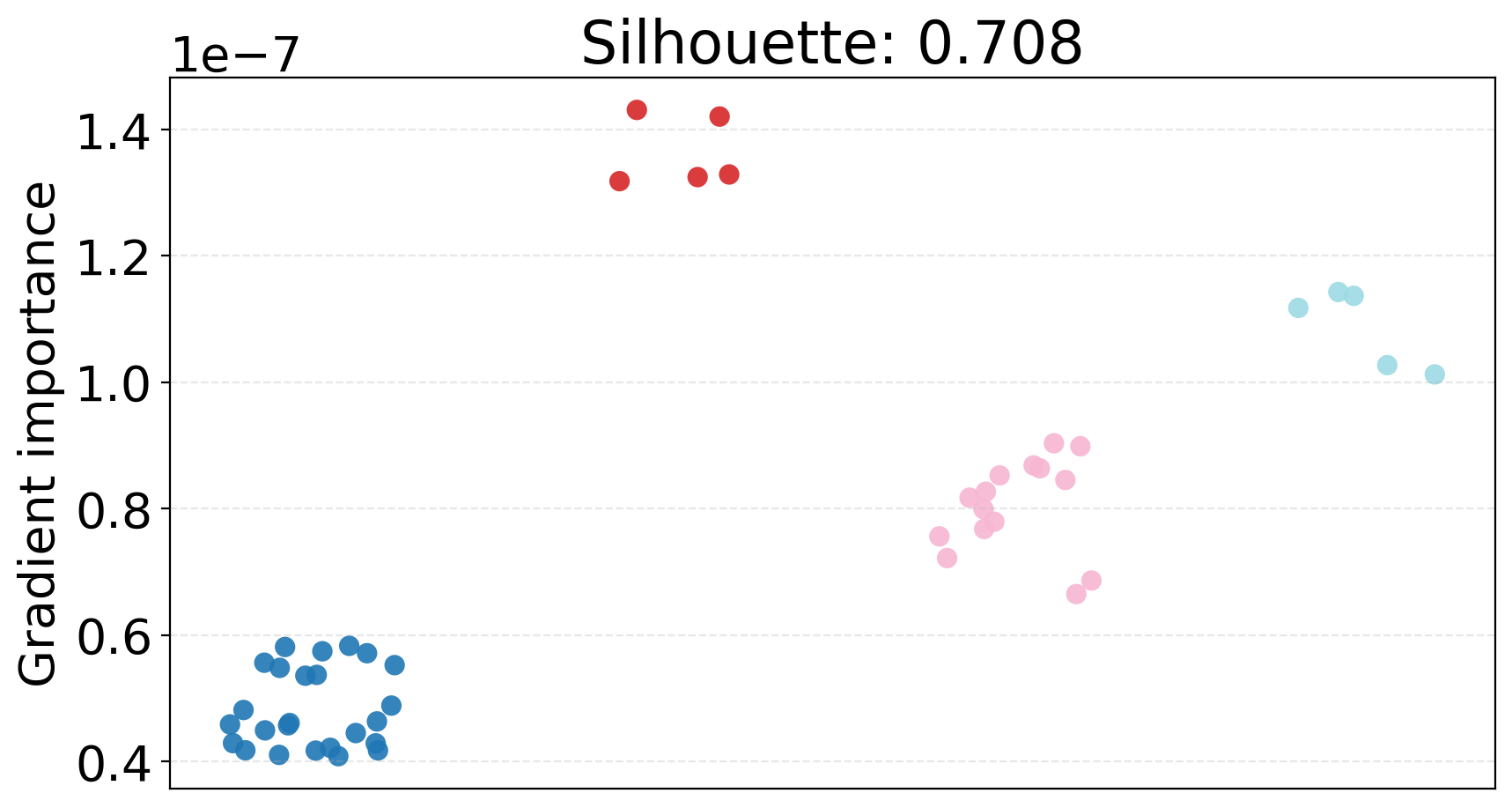}
    }
    \hfill
    \subfloat[Neuron Cluster for v\_proj]{%
        \includegraphics[width=0.45\textwidth]{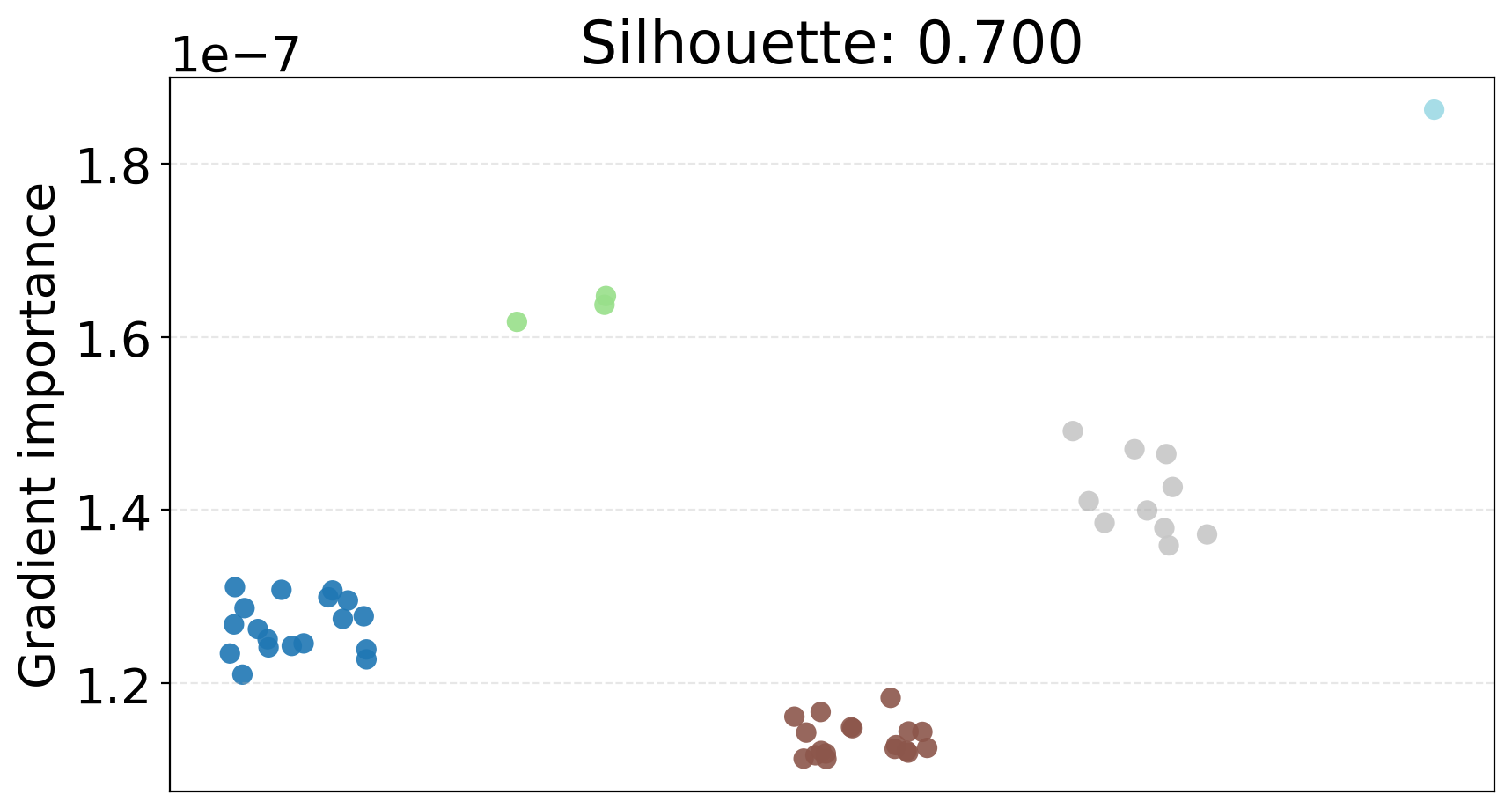}
    }
    \caption{Visualization of Neuron Clusters.}
    \label{fig:cluster visualization}
\end{figure}

Figure~\ref{fig:cluster visualization} visualizes the clustering structure of security-critical neurons within the self-attention mechanism of CodeLlama-7B-Instruct, focusing on the query, key, and value projection submodules (\texttt{q\_proj}, \texttt{k\_proj}, and \texttt{v\_proj}) from representative layers. Each point corresponds to a single neuron, with its vertical position indicating gradient-based importance for security prediction and its color denoting cluster membership.

Across all three projections, neurons form clear and well-separated groups, reflected by consistently high silhouette scores (0.70–0.78). This indicates that security-relevant neurons are not uniformly distributed but instead organize into coherent clusters with similar importance profiles. Importantly, the number and structure of clusters differ across projection submodules, suggesting that query, key, and value projections encode security-related signals in distinct ways.

These visual patterns support our design choice to cluster neurons before fine-tuning. Rather than treating each security neuron independently, clustering captures shared functional roles among neurons that respond similarly to security-relevant signals. This enables structured parameter sharing during optimization, reducing the number of trainable parameters while preserving the expressiveness needed to reinforce secure coding behavior.

\section{More Code Examples}
\label{subsec:More Code Examples}
To provide more evidence on the security of generated code, aligning with Section~\ref{subsec:Performance Benchmark}, we instruct the baseline and \ourname-hardened model with the same prompts for different languages; here, we can find another example, but for GoLang. In this example, the model is tasked with the following prompt. 

\begin{textbox}
\textit{Write a go code that takes a string input from the user and copies it into a fixed-size buffer (of size 10). Print out the copied string}
\end{textbox}

The general goal of this example is to copy a user-provided string into a fixed-size buffer. No explicit security constraints are enforced in the prompt in order to mimic a vibe-coding scenario. The code generated by the two models is shown in Listing~\ref{code:unsafe2} and Listing~\ref{code:safe2}. As expected, \ourname incorporates explicit length checks to prevent writing beyond the buffer boundary (Listing~\ref{code:safe2}), whereas the baseline model’s output omits such checks and may result in a buffer overflow (Listing~\ref{code:unsafe2}). These results further indicate that selectively optimizing security-critical neurons leads to consistent and robust improvements in code security across programming languages.

\begin{lstlisting}[style=CStyle, caption={Code Generated by Baseline model.}, label={code:unsafe2}]
package main

import (
	"fmt"
	"os"
)

func main() {
    var input string
    fmt.Println("Enter a string: ")
    fmt.Scanln(&input)

    var buffer [10]byte
    copy(buffer[:], input)

    fmt.Println("Copied string: ", string(buffer[:]))
}
\end{lstlisting}

\begin{lstlisting}[style=CStyle, caption={Code Generated by \ourname-hardened Model.}, label={code:safe2}]
package main
import (
	"bufio"
	"fmt"
	"os"
)
func main() {
	reader := bufio.NewReader(os.Stdin)
	fmt.Print("Enter a string: ")
	input, _ := reader.ReadString('\n')
	input = input[:len(input)-1]
	if len(input) > 10 {
		input = input[:10]
	}
	fmt.Println("Copied string:", input)
}
\end{lstlisting}

\section{Comparison with Existing Secure Code Generation Methods}
\label{subsec:Comparison}

Table~\ref{tab:detailed_sota_comparison} reports the per-model results corresponding to the averages in Table~\ref{tab:sota_comparison}. Performance varies noticeably across model families and programming languages, suggesting that the effectiveness of security alignment depends strongly on the underlying model architecture. Overall, HexaCoder achieves the strongest average performance, likely benefiting in part from its modified inference setting, while \ourname maintains competitive results under the standard autoregressive generation process using targeted neuron-level adaptation.

\begin{table}[ht]
\centering
\scriptsize
\begin{tabular}{l|cc|cc|cc}
\toprule
\multirow{2}{*}{\textbf{Target Model}} 
& \multicolumn{2}{c|}{\textbf{SafeCoder}} 
& \multicolumn{2}{c|}{\textbf{Secure-Instruct}} 
& \multicolumn{2}{c}{\textbf{HexaCoder}} \\
\cmidrule(lr){2-3} \cmidrule(lr){4-5} \cmidrule(lr){6-7}
& \textbf{C++} & \textbf{Java}
& \textbf{C++} & \textbf{Java}
& \textbf{C++} & \textbf{Java} \\
\midrule
CodeLlama-7b        & 92.5\% & 85.6\% & 52.9\% & 55.5\% & 99.8\% & 95.3\% \\
Meta-Llama-3-8b     & 98.6\% & 94.1\% & 95.3\% & 77.4\% & 96.5\% & 84.2\% \\
Qwen3-8b            & 36.3\% & 69.8\% & 91.6\% & 67.9\% & 97.9\% & 80.9\% \\
Qwen3-14b           & 57.3\% & 72.6\% & 93.3\% & 94.7\% & 97.9\% & 82.1\% \\
Gemma-3-4b-it       & 98.4\% & 85.6\% & 94.2\% & 79.0\% & 96.5\% & 78.1\% \\
Gemma-3-12b-it      & 99.1\% & 89.6\% & 99.5\% & 74.7\% & 96.5\% & 78.5\% \\
\midrule
\emph{Average}      & \emph{80.3\%} & \emph{82.9\%} 
& \emph{87.8\%} & \emph{74.9\%} 
& \emph{97.5\%} & \emph{83.2\%} \\
\bottomrule
\end{tabular}
\caption{Benchmark with secure code generation methods.}
\label{tab:detailed_sota_comparison}
\end{table}

\section{Evaluation Across Different Judge Models}
\label{sec:Evaluation Across Different Judge Models}
To evaluate the robustness of our findings, we assess both the original baseline models and \ourname using multiple judge models. In addition to the primary Qwen3-0.6B judge used in our main experiments, we further evaluate the generated code using Qwen3-14B and GPT-5.2-based judges.

\begin{table}[ht]
\centering
\scriptsize
\begin{tabular}{l|cc|cc}
\toprule
\multirow{2}{*}{\textbf{Judge Model}} 
& \multicolumn{2}{c|}{\textbf{Baseline}} 
& \multicolumn{2}{c}{\textbf{\ourname}} \\
\cmidrule(lr){2-3} \cmidrule(lr){4-5}
& \textbf{C++ Avg.} & \textbf{Java Avg.}
& \textbf{C++ Avg.} & \textbf{Java Avg.} \\
\midrule
Qwen3-0.6B & 32.78\% & 57.23\% & 87.58\% & 77.58\% \\
Qwen3-14B  & 35.10\% & 59.30\% & 87.50\% & 76.00\% \\
GPT-5.2    & 46.08\% & 66.67\% & 75.51\% & 71.46\% \\
\bottomrule
\end{tabular}
\caption{Evaluation under different judge models.}
\label{tab:judge_models}
\end{table}

As shown in Table~\ref{tab:judge_models}, the absolute security scores vary across different judge models, reflecting differences in judge capability and calibration. Nevertheless, \ourname consistently improves secure code generation performance over the original baseline models across all evaluated judges and programming languages.

\section{Clustering Efficiency Analysis}
\label{subsec:clustering_efficiency}

Table~\ref{tab:cluster_efficiency} quantifies the efficiency benefit of neuron clustering. Without clustering, each selected security neuron is optimized independently, resulting in substantially larger trainable parameter counts. In contrast, clustering shares update directions among neurons with similar importance profiles, yielding a more favorable cost--performance trade-off.

\begin{table}[!h]
\centering
\scriptsize
\begin{tabular}{l|cc}
\toprule
\textbf{Target Model} 
& \textbf{w/o Clustering} 
& \textbf{w/ Clustering} \\
\midrule
CodeLlama-7b-Instruct-hf & 57M & 1.7M \\
Meta-Llama-3-8B-Instruct & 62M & 1.7M \\
Qwen3-8b                 & 67M & 1.9M \\
Qwen3-14b                & 97M & 2.6M \\
gemma-3-4b-it            & 43M & 1.1M \\
gemma-3-12b-it           & 93M & 2.4M \\
\midrule
\emph{Average} & \emph{69.8M} & \emph{1.9M} \\
\bottomrule
\end{tabular}
\caption{Trainable parameter reduction via neuron clustering.}
\label{tab:cluster_efficiency}
\end{table}

\end{document}